\documentclass[5p,times]{elsarticle}

\usepackage{bm}        
\usepackage{amssymb}   
\usepackage{amsmath}
\usepackage{graphicx}  

\usepackage[squaren]{SIunits}
\usepackage{calc}
\usepackage{ifthen}

\usepackage{color}
\definecolor{grey}{rgb}{0.6,0.6,0.6}

\newcommand*{\Gr}[5]{%
  \makebox{%
    $#1#2#3\ifthenelse{\equal{#4}{}}{}{\!\times\!10^{#4}}\,\mathrm{#5}$%
  }%
}

\newcommand\abs[1]{\left|#1\right|}
\providecommand*{\vek}[1]{\text{\boldmath$#1$}} 
\newcommand\ind[1]{\textup{#1}}

\DeclareMathOperator{\der}{d\!}  
\DeclareMathOperator{\im}{\mathrm{i}}     

\newcounter{LaengeMasslinie}  
\newcounter{Laenge}
\newcounter{Hoehe}
\newcounter{MyBx}
\newcounter{MyBy}
\newcounter{MyCx}
\newcounter{MyCy}
\newcounter{MyDx}
\newcounter{MyDy}
\newcounter{MyEx}
\newcounter{MyEy}
\newcounter{MyFx}
\newcounter{MyFy}
\newcounter{MyGx}
\newcounter{MyGy}
\newcounter{MyHx}
\newcounter{MyHy}
\newcounter{MyIx}
\newcounter{MyIy}
\newcounter{MyJx}
\newcounter{MyJy}
\newcounter{MyKx}
\newcounter{MyKy}
\newcounter{MyLx}
\newcounter{MyLy}
\newcounter{MyMx}
\newcounter{MyMy}
\newcounter{LabAx}
\newcounter{LabAy}
\newcounter{LabBy}
\newlength{\BoxLaenge}       

\newcommand*{\Echo}[6]{%
  \setcounter{Laenge}{#2/2}
  \setcounter{Hoehe}{#2*\real{1.7}}
  \setcounter{MyBx}{#2*\real{0.25}}
  \setcounter{MyBy}{#2*\real{0.15}}
  \setcounter{MyCx}{#2*\real{0.3}}
  \setcounter{LabAx}{#2*\real{0.333}}
  \setcounter{LabAy}{#2*\real{0.7}}
  \setcounter{LabBy}{#2*\real{0.4}}
  \put(#1){
    \thicklines
    \qbezier(-\value{Laenge},0)(-\value{MyCx},0)(-\value{MyBx},\value{MyBy})
    \qbezier(-\value{MyBx},\value{MyBy})(0,\value{Hoehe})
            (\value{MyBx},\value{MyBy})
    \qbezier(\value{MyBx},\value{MyBy})(\value{MyCx},0)(\value{Laenge},0)
    \thinlines
    \put(#3){\put(\value{LabAx},\value{LabAy}){#4}}
    \put(#5){\put(0,-\value{LabBy}){\makebox[0mm]{#6}}}
  }
}

\newcommand*{\RFPuls}[5]{%
  \setcounter{LabAy}{#2*\real{1.05}}
  \put(#1){%
    \linethickness{#3}
    \put(0,0){\line(0,1){#2}}
    \put(#4){\put(0,\value{LabAy}){\makebox[0mm]{#5}}}
  }
}

\newcommand*{\Achse}[5]{%
  \put(#1){%
    \thinlines
    \ifthenelse{\equal{#2}{1,0}}{%
      \setcounter{LabAx}{#3-20}
      \put(0,0){\vector(#2){#3}}
      \put(#4){\put(\value{LabAx},-40){#5}}
    }{%
      \setcounter{LabAy}{#3-20}
      \put(0,0){\vector(#2){#3}}
      \put(#4){\put(-20,\value{LabAy}){\makebox[0mm][r]{#5}}}
    }
  }
}

\newcommand*{\MassPfeile}[6]{%
  \setlength{\BoxLaenge}{#3\unitlength}
  \setcounter{Laenge}{#3/2}
  \put(#1){
    \ifthenelse{\equal{#2}{1,0}}{%
      \put(\value{Laenge},#4){\vector(-1,0){\value{Laenge}}}
      \put(\value{Laenge},#4){\vector(1,0){\value{Laenge}}}
      \ifnum #4 < 0 {
        \setcounter{LaengeMasslinie}{10-#4}
        \multiput(0,0)(#3,0){2}{\line(0,-1){\value{LaengeMasslinie}}}
        \setcounter{LaengeMasslinie}{10+#4}
        \put(#5){\put(0,\value{LaengeMasslinie}){\makebox[\BoxLaenge]{#6}}}
        }
      \else {
        \ifnum #4 > 0 {
          \setcounter{LaengeMasslinie}{10+#4}
          \multiput(0,0)(#3,0){2}{\line(0,1){\value{LaengeMasslinie}}}
          \put(#5){\put(0,\value{LaengeMasslinie}){\makebox[\BoxLaenge]{#6}}}
        }
        \else {
          \setcounter{LaengeMasslinie}{10+#4}
          \put(#5){\put(0,\value{LaengeMasslinie}){\makebox[\BoxLaenge]{#6}}}
        }
        \fi
      }
      \fi
    }{%
      \put(#4,\value{Laenge}){\vector(0,-1){\value{Laenge}}}
      \put(#4,\value{Laenge}){\vector(0,1){\value{Laenge}}}
      \setcounter{LabAy}{#4-20}
      \put(#5){\put(\value{LabAy},0){
          \parbox[b][\BoxLaenge][c]{0mm}{
            \makebox[0mm][r]{#6}
          }
        }
      }
      \ifnum #4 < 0 {
        \setcounter{LaengeMasslinie}{10-#4}
        \multiput(0,0)(0,#3){2}{\line(-1,0){\value{LaengeMasslinie}}}
      }
      \else {
        \ifnum #4 > 0 {
          \setcounter{LaengeMasslinie}{10+#4}
          \multiput(0,0)(0,#3){2}{\line(1,0){\value{LaengeMasslinie}}}
        }
        \else {
        }
        \fi
      }
      \fi
    }
  }
}

\newcommand*{\PFGGrad}[7]{%
  \setcounter{Laenge}{#4/(#5+1)}
  \setcounter{LabAx}{#3*\real{1.1}}
  \put(#1){%
    \thinlines
    \ifthenelse{\equal{#2}{Pos}}{%
      \setcounter{LabAy}{#4*\real{0.9}}
      \thicklines
      \multiput(0,0)(#3,0){2}{\line(0,1){#4}} 
      \put(0,#4){\line(1,0){#3}}
      \thinlines
      \multiput(0,\value{Laenge})(0,\value{Laenge}){#5}{\line(1,0){#3}}
      \put(#6){\put(\value{LabAx},\value{LabAy}){\makebox[0mm][l]{#7}}}
    }{%
      \setcounter{LabAy}{#4*\real{1.1}}
      \thicklines
      \multiput(0,0)(#3,0){2}{\line(0,-1){#4}} 
      \put(0,-#4){\line(1,0){#3}}
      \thinlines
      \multiput(0,-\value{Laenge})(0,-\value{Laenge}){#5}{\line(1,0){#3}}
      \put(#6){\put(\value{LabAx},-\value{LabAy}){\makebox[0mm][l]{#7}}}
    }
  }
}

\newcommand*{\ImgEcho}[5]{%
  \setcounter{Laenge}{#3/2}
  \setcounter{Hoehe}{#2*\real{1.7}}
  \setcounter{MyBx}{#3*\real{0.025}}
  \setcounter{MyBy}{#2*\real{0.0}}

  \setcounter{MyCx}{#3*\real{0.03}}
  \setcounter{MyCy}{#2*\real{-0.4}}

  \setcounter{MyDx}{#3*\real{0.05}}
  \setcounter{MyDy}{#2*\real{0.0}}

  \setcounter{MyEx}{#3*\real{0.063}}
  \setcounter{MyEy}{#2*\real{0.25}}

  \setcounter{MyFx}{#3*\real{0.075}}
  \setcounter{MyFy}{#2*\real{0.0}}

  \setcounter{MyGx}{#3*\real{0.0825}}
  \setcounter{MyGy}{#2*\real{-0.12}}

  \setcounter{MyHx}{#3*\real{0.1}}
  \setcounter{MyHy}{#2*\real{0.0}}

  \setcounter{MyIx}{#3*\real{0.1126}}
  \setcounter{MyIy}{#2*\real{0.08}}

  \setcounter{MyJx}{#3*\real{0.155}}
  \setcounter{MyJy}{#2*\real{0.0}}

  \setcounter{MyKx}{#3*\real{0.175}}
  \setcounter{MyKy}{#2*\real{-0.05}}

  \setcounter{MyLx}{#3*\real{0.186}}
  \setcounter{MyLy}{#2*\real{-0.03}}

  \setcounter{MyMx}{#3*\real{0.192}}
  \setcounter{MyMy}{#2*\real{-0.01}}

  \setcounter{LabAx}{#3*\real{0.1}}
  \setcounter{LabAy}{#2*\real{0.7}}
  \put(#1){
   
    \qbezier(-\value{MyLx},\value{MyLy})
            (-\value{MyMx},\value{MyMy})
            (-\value{Laenge},0)
    \qbezier(-\value{MyJx},\value{MyJy})
            (-\value{MyKx},\value{MyKy})
            (-\value{MyLx},\value{MyLy})
    \qbezier(-\value{MyHx},\value{MyHy})
            (-\value{MyIx},\value{MyIy})
            (-\value{MyJx},\value{MyJy})
    \qbezier(-\value{MyFx},\value{MyFy})
            (-\value{MyGx},\value{MyGy})
            (-\value{MyHx},\value{MyHy})
    \qbezier(-\value{MyDx},\value{MyDy})
            (-\value{MyEx},\value{MyEy})
            (-\value{MyFx},\value{MyFy})
    \qbezier(-\value{MyBx},\value{MyBy})
            (-\value{MyCx},\value{MyCy})
            (-\value{MyDx},\value{MyDy})
    \qbezier(-\value{MyBx},\value{MyBy})
            (0,\value{Hoehe})
            (\value{MyBx},\value{MyBy})
    \qbezier(\value{MyBx},\value{MyBy})
            (\value{MyCx},\value{MyCy})
            (\value{MyDx},\value{MyDy})
    \qbezier(\value{MyDx},\value{MyDy})
            (\value{MyEx},\value{MyEy})
            (\value{MyFx},\value{MyFy})
    \qbezier(\value{MyFx},\value{MyFy})
            (\value{MyGx},\value{MyGy})
            (\value{MyHx},\value{MyHy})
    \qbezier(\value{MyHx},\value{MyHy})
            (\value{MyIx},\value{MyIy})
            (\value{MyJx},\value{MyJy})
    \qbezier(\value{MyJx},\value{MyJy})
            (\value{MyKx},\value{MyKy})
            (\value{MyLx},\value{MyLy})
    \qbezier(\value{MyLx},\value{MyLy})
            (\value{MyMx},\value{MyMy})
            (\value{Laenge},0)
    \thinlines
    \put(#4){\put(\value{LabAx},\value{LabAy}){#5}}
  }
}

\begin{document}

\begin{frontmatter}

\title{Phase Incremented Echo Train Acquisition applied to Magnetic Resonance Pore Imaging}

\author[vic]{S.A. Hertel\fnref{fn1}}
\ead{stefan.hertel@shell.com}
\fntext[fn1]{Currently at Shell Intl. E\&P Inc., Rock and Fluid Physics, Houston, TX}

\author[vic]{P. Galvosas}

\address[vic]{MacDiarmid Institute for Advanced Materials and
  Nanotechnology, School of Chemical and Physical Sciences, Victoria
  University of Wellington, Wellington 6140, New Zealand}

\begin{abstract}
  Efficient phase cycling schemes remain a challenge for NMR
  techniques if the pulse sequences involve a large number of
  rf-pulses. Especially complex is the Carr Purcell Meiboom Gill
  (CPMG) pulse sequence where the number of rf-pulses can range from
  hundreds to several thousands. Our recent implementation of Magnetic
  Resonance Pore Imaging (MRPI) is based on a CPMG rf-pulse sequence
  in order to refocus the effect of internal gradients inherent in
  porous media. While the spin dynamics for spin-$1/2$ systems in CPMG
  like experiments are well understood it is still not straight
  forward to separate the desired pathway from the spectrum of
  unwanted coherence pathways. In this contribution we apply Phase
  Incremented Echo Train Acquisition (PIETA) to MRPI. We show how
  PIETA offers a convenient way to implement a working phase cycling
  scheme and how it allows one to gain deeper insights into the
  amplitudes of undesired pathways.
\end{abstract}

\end{frontmatter}

\section{Introduction}

The motion of spin bearing particles can be studied by the judicial
use of pulsed magnetic field gradients in NMR
experiments~\cite{Stejskal65JCP_I,Callaghan11Book}. Pulsed Gradient
Spin Echo (PGSE) NMR is best known for diffusion measurements in
porous media as well as for velocimetry
studies~\cite{Callaghan11Book}. The potential of PGSE NMR for NMR
microscopy has been demonstrated with the appearance of diffraction
patterns when self-diffusion was studied in a bead pack
sample~\cite{Callaghan91Nature}. Magnetic Resonance Pore Imaging
(MRPI) is a recent advancement in PGSE NMR which dramatically
increases the resolution as compared to MRI for suitable closed pore
systems~\cite{Laun11PhysRevLett}. MRPI relies on a weak gradient pulse
which can be several hundreds of milliseconds long. During this
gradient pulse the nuclear spins have to be kept in the transverse
plane~\cite{Laun12PhysRevE,Kuder13PhysRevLett}. We suggested therefore
a CPMG based design in order to refocus the internal gradients and
other experimental imperfections~\cite{Hertel13PhysRevE,
  Hertel15PhysRevE}. However, the introduction of many rf-pulses in
the presence of magnetic field gradients leads to a complicated
coherence pathway spectrum~\cite{Goelman95JMR, Huerlimann01JMR}. This
may not always be a problem in CPMG relaxation studies, since the
decay due to transverse relaxation is still
obtained~\cite{Song02JMR}. In MRPI however, it is essential to
preserve the spin phase throughout the rf-pulse train, while the
unwanted pathways do interfere
adversely~\cite{Hertel15PhysRevE}. Since traditional phase cycling
techniques proved impractical for our CPMG pulse sequence design, it
was necessary to explore alternative ways to filter for the correctly
encoded signal. This study focuses on the application of a novel phase
cycling technique called Phase Incremented Echo Train Acquisition
(PIETA)~\cite{Baltisberger12JChemPhys}.

In pulsed NMR the spectrum of coherence pathways (CPs) is multiplied
with each additional rf-pulse~\cite{Levitt01Book}. This creates a very
rich spectrum of CPs, where each CP contains specific information
which is influenced by its history since excitation. Since each
coherence may experience different physical interactions, it is
paramount to select the right CPs of the spin system under study. The
most popular strategy to achieve this selection involves the nested
cycling of the rf-pulse phases and the receiver phase from scan to
scan according to a master
equation~\cite{Bain84JMR,Bodenhausen84JMR}. However, the number of CPs
grows exponentially with the number of rf-pulses and finding the right
phase cycling scheme may become complicated.  Throughout the article
we consider a spin-$1/2$ system without coupling between the
individual spins and where the desired CP does not involve any
$z$-storage. In this case, an NMR pulse sequence with $10$ rf-pulses
would need $N=3^{10}=59049$ phase steps following the nested phase
cycling scheme~\cite{Hughes04JMR}. Several algorithms were designed to
solve this problem with the help of computer
programs~\cite{Jerschow98JMR,Ollerenshaw00JMR,McClung99ConMagRes}.
Two other widely applied strategies are improved perfection of the
spin rotations induced by specially designed
rf-pulses~\cite{Levitt79JMR} and the application of pulsed field
gradients~\cite{Bax80ChemPhysLett}. These two latter strategies can
reduce the complexity of the problem, but they are often harder to
implement or face limitations due to the available
hardware. Especially the Carr-Purcell-Meiboom-Gill (CPMG) NMR pulse
sequence~\cite{Carr54PhysRev,Meiboom58RSI} and its derivatives can
contain several hundreds or even thousands of rf-pulses and the number
of phase steps with traditional phase cycling would be forbiddingly
high. A first solution emerged with the development of Cogwheel phase
cycling which when applied to CPMG like pulse trains promised
significant reductions in the number of phase steps
needed~\cite{Levitt02JMR}. When Cogwheel is applied to CPMG the
receiver phases have to be cycled such that they span a $2\pi$ angle
range after $n_{\textup{rf}}$ acquisitions, where $n_{\textup{rf}}$ is
the number of $180\degree$ rf-pulses~\cite{Levitt02JMR}.  This renders
this method appealing for experiments with a fixed number of
rf-pulses. However, in the case of CPMG echo trains with a variable
number of rf-pulses it would require as many independent measurements
(including the Cogwheel phase cycle) as different numbers of rf-pulses
have been chosen. A recent approach called Phase Incremented Echo
Train Acquisition (PIETA) offers the extraction of all echoes in a
given rf-pulse train and a de-convolution of the contributing
coherence pathways with the same number of scans as
Cogwheel~\cite{Baltisberger12JChemPhys}.

In this work, PIETA is utilized to investigate the coherence pathway
spectrum contributing to the spin echo signal of MRPI. Furthermore, it
is shown how the artifact free coherence pathway can be selected.

\section{Theory}

\subsection{Phase cycling}

In NMR experiments the desired information is usually contained in one
specific coherence pathway (CP)
$\bm{p}=(p_{\text{0}},p_{\text{1}},p_{\text{2}},\allowbreak\dots ,p_{\text{n}})$ or
a small subset $\mathcal{P}$ of all CPs~\cite{Ernst87Book}. The values
$p_{\textup{k}}$ denote the coherence order of the spins after the
$k$-th rf-pulse of the NMR pulse sequence. The elements
$p_{\textup{k}}$ can assume integer values which are determined by the
spin quantum number $I$ and the coupling between multiple spins. For a
system of uncoupled spin-$1/2$ nuclei the values of $p_{\textup{k}}$
can be $+1,-1$ and $0$. It is assumed that every CP starts from a well
defined equilibrium with $p_{\textup{0}}=0$ (associated with
magnetization in $z$-direction) before the first rf-pulse. For a CP to
be measured with quadrature NMR detection it has to end with
$p_{\textup{n}}=-1$~\cite{Ernst87Book}. The magnetization of all three
levels $\{+1,-1,0\}$ is mixed with each additional rf-pulse giving
rise to a large number of CPs growing exponentially with $3^n$, where
$n$ is the number of rf-pulses~\cite{Song02JMR}. The complex NMR
signal for a single scan with scan counter $m$ can be expressed as
\begin{equation}
\label{eq:signal}
s_{\textup{m}}(t)=\sum_{\vek{p}}s_{\vek{p}}(t)\exp\{-\im\phi_{\vek{p}}(m)\}\,,
\end{equation}
where $s_{\vek{p}}(t)$ is the signal arising from a specific coherence
pathway $\vek{p}$ after the last rf-pulse. $\phi_{\vek{p}}(m)$ denotes
the final phase with which the CP is recorded in the spectrometer
memory and it is given by~\cite{Levitt02JMR}
\begin{equation}
\label{eq:phase}
\phi_{\vek{p}}(m)=\Delta p_{\textup{1}}\phi_{\textup{1}}(m)+\Delta p_{\textup{2}}\phi_{\textup{2}}(m)+\dots +\Delta p_{\textup{n}}\phi_{\textup{n}}(m)+\phi_{\textup{rec}}(m)\,.
\end{equation}
Here, $\phi_{\textup{k}}$ are the phases of the $k$-th rf-pulse and
$\Delta p_{\textup{k}}=p_{\textup{k}}-p_{\textup{k}-1}$ is the
coherence transfer difference induced with the $k$-th rf-pulse. The
signal after $N$ scans can be expressed as the complex sum given by

\begin{align}
  \label{eq:signal_all_scans}
s(t)&=\sum_{m=1}^{N}s_{\ind{m}}(t)\\
&=\sum_{m=1}^{N}\sum_{\vek{p}}s_{\vek{p}}(t)\exp\{-\im\phi_{\vek{p}}(m)\}\nonumber\\
&=\sum_{\vek{p}}s_{\vek{p}}(t)\sum_{m=1}^{N}\exp\{-\im\phi_{\vek{p}}(m)\}\,,\nonumber
\end{align}
where we inserted eq.~\ref{eq:signal} in the first step and in the
second step we used the fact that $s_{\vek{p}}(t)$ does not depend on
$m$. The last term in eq.~\ref{eq:signal_all_scans} shows that each
individual CP signal is multiplied by the sum of phase factors added
over all $N$ scans. Thus, in order to filter any desired pathway
$\vek{p}$ after $N$ scans, the following condition has to be
full-filled
\begin{equation}
  \label{eq:condition}
  \sum_{m=1}^{N}\exp\{-\im\phi_{\vek{p}}(m)\}=\begin{cases}
    N & \text{if}\,\,\,\, \vek{p}\in \mathcal{P}\\
    0 & \text{otherwise}\,.\\
  \end{cases}
\end{equation}
The most popular strategies to achieve this condition are the nested
phase cycling approach~\cite{Bain84JMR,Bodenhausen84JMR} and Cogwheel
phase cycling scheme~\cite{Levitt02JMR}. Both methods involve the
design of a set of $\phi_{\textup{k}}(m)$ for each scan $m$ and to
cycle the receiver phase $\phi_{\textup{rec}}$ such that the wanted CP
is always recorded with the same phase, while all other CPs cycle in
phase and cancel upon completion of the full set of scans.  In the
case of many rf-pulses, such as they are common in CPMG type
experiments, these methods are hard to apply because of the shear
number of CPs and they do not provide any extra information about the
importance of the unwanted CPs.

A different concept was proposed with Multiplex phase
cycling~\cite{Ivchenko03JMR} where each scan is recorded separately in
the spectrometer memory and the wanted CPs are selected in a
post-processing step. It will be shown in the next section how PIETA
is similar to both multiplexing and Cogwheel phase cycling.

\subsection{Phase Incremented Echo Train Acquisition}
\label{sec:PIETA}
Baltisberger et al.~\cite{Baltisberger12JChemPhys} introduced a new
phase cycling scheme which allows one to extract the direct CP from
Echo Train Acquisition (ETA) experiments. The authors use ETA as a
general expression for a train of spin echos due to a train of
rf-pulses, which can be appended to an NMR experiment or may represent
a stand-alone experiment like CPMG. This novel approach was called
Phase Incremented Echo Train Acquisition (PIETA), because the phase of
every other refocusing pulse is incremented as a single variable
$\phi_{\textup{P}}$ from scan to scan.\label{pp:one_echo} Note that
the application of PIETA in this contribution slightly deviates from
the use cases suggested by Baltisberger
\emph{et. al}~\cite{Baltisberger12JChemPhys}. Instead of recording
many spin echo amplitudes for each scan, here only one spin echo is
recorded at the end of the rf-pulse train for each scan as will be
explained below.
\begin{figure}[htb]
  \centering
  \setlength{\unitlength}{0.07mm} 
  \begin{picture}(1200,700)(-110,-350)
    \Achse{0,50}{1,0}{1100}{0,0}{$t$}
  \RFPuls{0,50}{200}{.75mm}{3,13}{$90^{\circ}_{x}$}
  \MassPfeile{0,0}{1,0}{200}{-20}{0,-60}{$\tau$}
  \put(0,0){\line(0,1){50}}
  \put(200,0){\line(0,1){50}}
  \RFPuls{200,50}{200}{1.5mm}{5,15}{$180^{\circ}_{\phi_{\text{P}}}$}
  \MassPfeile{200,0}{1,0}{400}{-20}{0,-60}{$2\tau$}
  \ImgEcho{400,50}{100}{300}{0,0}{$\phi_{\textup{rec}}$}
  \RFPuls{600,50}{200}{1.5mm}{5,15}{$180^{\circ}_{\frac{\pi}{2}}$}
  \ImgEcho{800,50}{100}{300}{0,0}{$\phi_{\textup{rec}}$}
  \MassPfeile{600,0}{1,0}{400}{-20}{0,-60}{$2\tau$}
  \put(600,0){\line(0,1){50}}
  \put(180,-350){\dashbox{10}(820,600)[]{}}
  \put(920,-380){\makebox(30,-5)[l]{loop}}
  \put(-90,-225){\makebox(30,-5)[l]{$p$}}
  \put(-45,-225){\makebox(30,-5)[r]{$0$}}
  \put(-45,-150){\makebox(30,-5)[r]{$+1$}}
  \put(-45,-300){\makebox(30,-5)[r]{$-1$}}

  {\color{grey}{%
  \put(0,-150){\line(1,0){1100}}
  \put(0,-225){\line(1,0){1100}}
  \put(0,-300){\line(1,0){1100}}
  }}


\linethickness{0.4mm}

  \put(0,-225){\line(0,-1){75}}
  \put(0,-300){\line(1,0){185}}
  \put(185,-300){\line(1,6){25}}
  \put(210,-150){\line(1,0){375}}
  \put(585,-150){\line(1,-6){25}}
  \put(610,-300){\line(1,0){390}}
  \multiput(0,-225)(0,15){5}%
  {\line(0,1){10}}
  \multiput(0,-150)(15,0){13}%
  {\line(1,0){10}}
  \multiput(187,-160)(2,-10){15}%
  {\line(0,1){7}}
  \multiput(215,-300)(15,0){25}%
  {\line(1,0){10}}
  \multiput(587,-290)(2,10){15}%
  {\line(0,-1){7}}
  \multiput(615,-150)(15,0){26}%
  {\line(1,0){10}}
  \end{picture}
  \caption{PIETA phase scheme as applied to the CPMG pulse sequence,
    where every second $180^{\circ}$ rf-pulse is phase incremented
    with the rf-pulse phase $\phi_{\textup{P}}$. The phases of the
    $90^{\circ}$ rf-pulse and every even $180^{\circ}$ rf-pulse as
    well as the receiver phase $\phi_{\textup{rec}}$ stay fixed.}
  \label{fig:PIETA_sequence}
\end{figure}
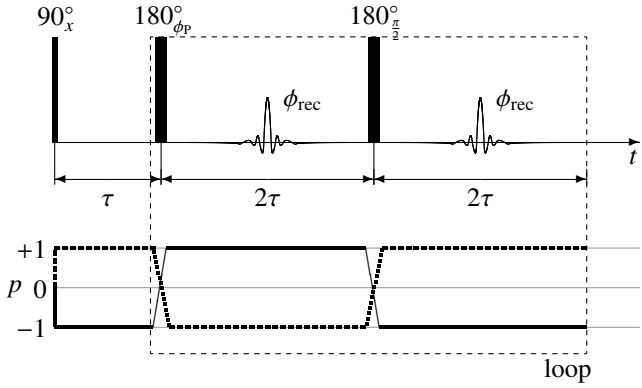
PIETA applied to the CPMG pulse sequence is shown in
fig.~\ref{fig:PIETA_sequence}. The part enclosed by the dashed line is
repeated multiple times to create a train of refocusing $180\degree$
rf-pulses. The rf-pulse phase $\phi_{\textup{P}}$ is incremented from
scan to scan, while the phases of the leading $90\degree$ rf-pulse, of
every other $180\degree$ rf-pulse and of the receiver phase
$\phi_{\textup{rec}}$ stay fixed for all scans. The level diagram
below the rf-pulses in fig.~\ref{fig:PIETA_sequence} shows the three
coherence orders $0,+1,-1$. In this level diagram the direct CP which
is measured for even echos $\vek{p}=(0,-1,+1,-1,\dots,+1,-1)$ is
indicated as a solid line, while its conjugate for the odd echos
$\vek{\tilde{p}}=(0,+1,-1,+1,-1\dots,+1,-1)$ is shown as a dashed
line. With PIETA the vector of spectrometer phases for the $m$-th scan
is given by
$\bm{\phi}(m)=(\phi_{\textup{1}}=0,\allowbreak\phi_{\textup{P}}=\nu
m,\allowbreak\phi_{\textup{3}}=\pi/2,\phi_{\textup{P}}=\nu
m,\dots,\phi_{\textup{rec}}=0)$, where $\nu$ is a winding number given
by
\begin{equation}
  \label{eq:nu}
  \nu=\frac{2\pi}{N}\,.
\end{equation}
Thus, the phase $\phi_{\textup{P}}$ is rotated by $2\pi$ when all $N$
scans are completed. The final signal phase for the PIETA experiment
as shown in fig.~\ref{fig:PIETA_sequence} is given by
\begin{equation}
  \label{eq:phase_pieta}
  \phi_{\vek{p}}(m)=\nu m \Delta p_{\textup{2}}+\nu m \Delta p_{\textup{4}}+\dots\\
\end{equation}
where the phases which remain fixed for all scans have been
neglected. The fixed phases lead to a constant phase for all scans and
will not be needed in the following analysis. One may write
eq.~\ref{eq:phase_pieta} as a sum leading to
\begin{align}
  \label{eq:deltaP}
  \phi_{\vek{p}}(m)&=\nu m \sum_{\textup{k even}}\Delta p_{\textup{k}}\\
  &=\nu m \Delta P\,,\nonumber
\end{align}
where $\Delta P$ is defined as the cumulative coherence transfer
difference induced by the even rf-pulses 
\label{pp:delta_p_def} $\left(\Delta P=\sum_{\textup{k even}}\Delta
  p_{\textup{k}}\right)$ which are phase cycled with
$\phi_{\text{P}}$. Therefore, the NMR signal as given by
eq.~\ref{eq:signal} can be rewritten according to
\begin{equation}
  \label{eq:pieta_signal}
  s_{\textup{m}}(t)=\sum_{\vek{p}}s_{\vek{p}}(t)\exp\{-\im\nu m \Delta P\}\,.
\end{equation}
It can be seen that $\Delta P$ assumes the role of a modulation
frequency when the signal is acquired as a function of the phase
$\phi_{\textup{P}}=\nu m$. 

Similar to the multiplex phase cycling scheme~\cite{Ivchenko03JMR},
all the spin echos in the CPMG train are stored separately for each
scan in the computer memory, while $\phi_{\textup{P}}$ is incremented
from scan to scan. Let the number of spin echos recorded be denoted by
$n_{\text{E}}$ then the stored data is the signal matrix
$s(n_{\text{E}},t,\phi_{\textup{P}})$. In this matrix, $t$ represents
the discrete time steps which start all over again with each spin echo
$n_{\text{E}}$. One may recall that $\phi_{\textup{P}}$ is connected
with $\Delta P$ as the underlying modulation frequency (see
eq.~\ref{eq:deltaP}) which implies that $s(n_{\text{E}},t,\Delta P)$
and $s(n_{\text{E}},t,\phi_{\textup{P}})$ are a Fourier pair similar
to the relation found in~\cite{Drobny78SympFaradSoc}. Thus, after all
scans are completed one can perform a discrete Fourier transform in
the $\phi_{\textup{P}}$ direction of the data matrix to obtain the
signal in the $\Delta P$ dimension according to
\begin{align}
  \label{eq:FT}
  s(n_{\text{E}},t,\Delta P)&=\mathcal{F}\{s(n_{\text{E}},t,\phi_{\textup{P}})\}\\
 &= \int s(n_{\text{E}},t,\phi_{\textup{P}}) \exp\{-\im\phi_{\textup{P}}\Delta P \}\der\phi_{\textup{P}}\,.\nonumber
\end{align}
The crucial observation of Baltisberger et
al.\cite{Baltisberger12JChemPhys} is that $\Delta P$ is unique for the
direct CP and can be extracted from $s(n_{\text{E}},t,\Delta P)$
without interference of any other CPs.

\label{pp:comment_9} For the isolated spin-$1/2$ system considered
here, the direct CP difference vector for even $n_{\text{E}}$ is given
by $\Delta \bm{p}=(-1,+2,-2,\allowbreak\dots\allowbreak ,+2)$ and for
odd $n_{\text{E}}$ is given by $\Delta
\tilde{\bm{p}}=(+1,-2,+2,\allowbreak\dots\allowbreak ,-2)$. The CP
difference for each rf-pulse cycled with $\phi_{\textup{P}}$ is equal
to $\Delta p_{\textup{k}}=+2$, while for odd spin echos it is equal to
$\Delta \tilde{p}_{\textup{k}} =-2$ since the conjugate CP
$\tilde{\bm{p}}$ is measured. All other CP vectors have one or more
elements with $\abs{\Delta p_{\textup{k}}}<2$ and thus they have a
smaller cumulative coherence pathway difference $\Delta P$.

$\Delta P$ for the direct coherence pathway in dependence of the spin
echo count is given by~\cite{Walder13JCP}
\begin{equation}
\label{eq:direct_selection}
\Delta P = (-1)^{n_{\text{E}}}\left\{2\left\lfloor\frac{n_{\text{E}}-1}{2}\right\rfloor+2\right\}\,,
\end{equation}
where $\left\lfloor x\right\rfloor$ is the integer floor
function. Note that for CPMG the spin echo count as a function of the
number of rf-pulses $n$ is given by $n_{\text{E}}=n-1$ in order to account
for the $90\degree$ excitation pulse.

There is a minimum number of scans needed in order to completely
de-aliase the direct CP, since the number of phase increments has to
be chosen such that the Nyquist-Shannon criterion
\cite{Shannon49ProcInstRadEng} is full-filled. 
In the case of PIETA the sampling
frequency is given by $\nu=2\pi/N$. This frequency has to be at least
double the frequency which shall be detected. The highest frequency to
be detected is is given by $\Delta P(n_{\text{E}})$ for the direct CP
as given by eq.~\ref{eq:direct_selection}. Thus, if the spin echo
count is even then $\Delta P=n_{\text{E}}$ and for odd spin echoes it
is $\Delta P=-(n_{\text{E}}+1)$. Therefore, the number of phase
increments in dependence on the echo count is given by
\begin{equation}
  \label{eq:Nscan_pieta}
  N\geq\begin{cases}
    2n_{\text{E}} & \text{if}\,\,n_{\text{E}}\,\,\text{even}\\
    2(n_{\text{E}}+1) & \text{if}\,\,n_{\text{E}}\,\,\text{odd}\,.\\
  \end{cases}
\end{equation}

\label{pp:one_echo_II} In this work, the spin echo count $n_{\textup{E}}$
will not be increased as compared to conventional CPMG relaxation
experiments. Instead, $n_{\textup{E}}$ stays fixed and the gradient
amplitude is stepped as will be explained in the next section.


\subsection{Magnetic Resonance Pore Imaging}

Magnetic Resonance Pore Imaging (MRPI) can be regarded as a Pulsed
Gradient Spin Echo (PGSE) NMR technique. PGSE NMR techniques measure
the NMR signal as a function of the wave vector $\vek{q}=\gamma\delta
\vek{G}/2\pi$, where $\vek{G}$ denotes the gradient amplitude,
$\gamma$ is the gyro-magnetic ratio of the observed nucleus and
$\delta$ is the length of the gradient pulses. The PGSE NMR signal is
usually normalized with the NMR signal obtained with zero gradient
pulse amplitude $E(q)=M(\vek{q})/M(\vek{q}=0)$. Recently, Laun
\textit{et al.}~\cite{Laun11PhysRevLett} suggested the gradient pulse
pattern used for MRPI, which created a hybrid between MRI and PGSE
NMR. With MRPI the spin echo attenuation is resembling the averaged
form factor $\overline{S_{\textup{0}}(\vek{q})}$ of the pores as
measured in the direction of $\vek{q}$. The form factor is the Fourier
transform of the pore shape $\rho_{\textup{0}}(\vek{r})$. Therefore,
one may obtain an average image of the pores by performing an inverse
Fourier transformation of the MRPI spin echo attenuation if certain
requirements are met, i.e.
$\overline{\rho_{\textup{0}}(\vek{r})}=\mathcal{F}^{-1}\{\overline{S_{\textup{0}}(\vek{q})}\}$.
Note that the pore shape $\rho_{\textup{0}}(\vek{r})$ equals the spin
density distribution if the spin density is homogeneous throughout the
pore space. In this contribution we discuss the experimental results
for a cylindrical capillary sample for which the form factor is given
by
\begin{equation}
  \label{eq:cyl_formfactor}
  S_{\text{0}}(q)=\frac{J_{\textup{1}}(2\pi qL)}{\pi q L}\,,
\end{equation}
where $J_{\textup{1}}$ denotes the cylindrical Bessel function of
first order and $L$ is the radius of the cylindrical domain.  
For obtaining eq.~\ref{eq:cyl_formfactor} the wave-vector $\vek{q}$
has to be applied perpendicular to the cylinder axis. In the following
discussion it will be assumed that the gradient pulses will always be
applied in this direction and therefore the vector notation is
dropped.

\begin{figure}[htb]
  \centering
  \setlength{\unitlength}{0.08mm}
  \begin{picture}(900,250)(-100,-170)
  \Achse{0,0}{1,0}{850}{0,0}{$t$}%
  \Achse{0,-10}{0,1}{80}{10,-20}{\makebox[0mm][r]{$G^*$}}%
  \put(30,0){%
    \PFGGrad{10,0}{Neg}{580}{20}{0}{-90,20}{}%
    \PFGGrad{610,0}{Pos}{100}{70}{0}{-90,20}{}%
    \MassPfeile{10,0}{1,0}{580}{-65}{0,3}{$\delta_\text{L}$}%
    \MassPfeile{610,0}{1,0}{100}{-65}{0,3}{$\delta_\text{N}$}%
    \MassPfeile{10,0}{1,0}{700}{-120}{0,0}{$T$}%
  }%
  \put(530,20){\vector(0,-1){20}}
  \put(530,-40){\vector(0,1){20}}
  \put(535,20){$G_{\text{L}}$}
  \MassPfeile{735,0}{0,1}{70}{10}{75,0}{$G_{\text{N}}$}
  \put(-80,-10){\makebox[0mm][c]{\textbf{(a)}}}%
\end{picture}\\
\begin{picture}(900,250)(-100,-110)
 \Achse{0,0}{1,0}{850}{0,0}{$t$}%
 \Achse{0,-10}{0,1}{80}{10,-20}{\makebox[0mm][r]{$G\,$}}%
 \color{grey}
   \put(0,10){\line(1,0){850}}
 \color{black}
\put(30,0){%
 \multiput(10,0)(120,0){5}{\PFGGrad{0,0}{Pos}{40}{20}{0}{-90,20}{}}%
 \multiput(70,0)(120,0){5}{\PFGGrad{0,0}{Neg}{40}{20}{0}{-90,20}{}}%
 \PFGGrad{610,0}{Neg}{40}{70}{0}{-90,20}{}
 \PFGGrad{670,0}{Pos}{40}{70}{0}{-90,20}{}
 \multiput(60,0)(60,0){12}{\RFPuls{0,0}{100}{.3mm}{5,15}{}}
 }%
 \RFPuls{30,0}{100}{.15mm}{5,10}{\small{$90\degree$}}
 \RFPuls{690,0}{100}{.2mm}{5,12}{\small{$180\degree$ pulses}}

  \put(770,30){\vector(0,-1){20}}
  \put(770,-20){\vector(0,1){20}}
  \put(780,35){$g_{\textup{int}}$}
  \put(-80,-10){\makebox[0mm][c]{\textbf{(b)}}}%
\end{picture}
\begin{picture}(900,150)(-100,-70)
  \Achse{0,0}{1,0}{850}{0,0}{$t$}%
  \Achse{0,-10}{0,1}{80}{10,-20}{\makebox[0mm][r]{$G^*$}}%
  \put(30,0){%
    \multiput(10,0)(60,0){10}{\PFGGrad{0,0}{Neg}{40}{20}{0}{-90,20}{}}%
    \multiput(610,0)(60,0){2}{\PFGGrad{0,0}{Pos}{40}{70}{0}{-90,20}{}}%
 \color{grey}
 \multiput(60,0)(120,0){6}{\PFGGrad{0,0}{Neg}{60}{10}{0}{-90,20}{}}%
 \multiput(120,0)(120,0){5}{\PFGGrad{0,0}{Pos}{60}{10}{0}{-90,20}{}}%
 \color{black}    
  }%
 \color{grey}
 \PFGGrad{30,0}{Pos}{60}{10}{0}{-90,20}{}%

 \put(750,10){\line(0,-1){10}}
 \put(750,10){\line(1,0){100}}

 \color{black}    
  \MassPfeile{460,25}{1,0}{40}{-10}{10,10}{$\delta^{'}$}
  \put(475,-30){\vector(1,0){25}}
  \put(545,-30){\vector(-1,0){25}}
  \put(500,-35){\line(0,1){10}}
  \put(520,-35){\line(0,1){10}}
  \put(500,-70){$\delta_{\textup{s}}$}
  \put(-80,-10){\makebox[0mm][c]{\textbf{(c)}}}%
\end{picture}%
\caption{Gradient pulse pattern of the MRPI pulse sequence as proposed
  by Laun et. al~\cite{Laun11PhysRevLett} (a). The presence of a
  constant internal gradient $g_{\textup{int}}$ (grey) required the
  replacement used for this work which consisted of a sequence of
  gradient pulses interspersed with $180\degree$ rf-pulses (b). The
  pulsed gradients effectively add up, while the internal gradient
  $g_{\textup{int}}$ is compensated (c).}
\label{fig:mrpi_vs_pgse}
\end{figure}
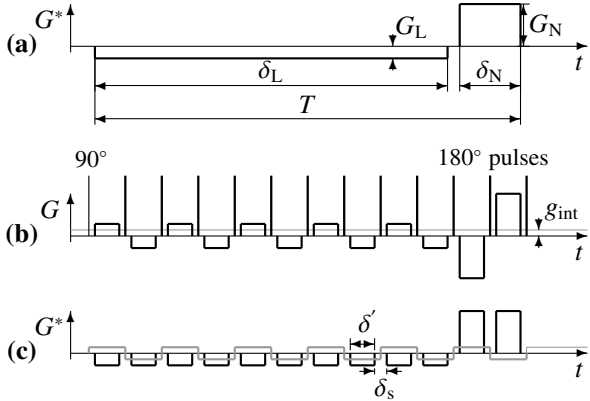

MRPI relies on the application of a long gradient pulse with duration
$\delta_{\textup{L}}$ and amplitude $G_{\textup{L}}$ as shown in
fig.~\ref{fig:mrpi_vs_pgse} (a). The long gradient pulse allows for
diffusion encoding, while the intense gradient with amplitude
$G_{\textup{N}}$ and duration $\delta_{\textup{N}}$ acts similar to a
phase imaging gradient in MRI~\cite{Laun11PhysRevLett}. These two
gradient pulses have to full-fill the spin echo condition
$\int_{0}^{T}G^*(t)=0$ which requires
$G_{\textup{L}}\delta_{\textup{L}}=G_{\textup{N}}\delta_{\textup{N}}$.
\label{pp:def_of_T} Here, $T$ is defined as the time from the onset of
the first gradient pulse until the the end of the last gradient pulse.
Fig.~\ref{fig:mrpi_vs_pgse} (b) shows schematically the replacement of
the gradient pulse pattern as initially suggested
in~\cite{Laun11PhysRevLett} with the one used in this work. The
gradient pulses were applied with alternating polarity in the
laboratory reference frame while they were interspersed with
$180\degree$ rf-pulses. The replacement was chosen to minimize the
dephasing due to the internal gradients $g_{\textup{int}}$ by
refocusing their influence on the spin
phases~\cite{Hertel13PhysRevE}. Fig.~\ref{fig:mrpi_vs_pgse} (c) shows
the effective gradient pattern $G^*(t)$ as experienced by the spin
system when the effect of the $180\degree$ pulses is accounted
for. The pulsed gradients are effectively adding up to yield a pattern
similar to fig.~\ref{fig:mrpi_vs_pgse} (a), while the internal
gradients are compensated. Note that many more gradient pulses were
used in the actual implementation, which will be shown in
sec.~\ref{sec:exp} along with the rf-pulse phases.

For illustration purposes, consider the gradient pattern as shown in
fig.~\ref{fig:mrpi_vs_pgse} (a). In the ideal case that the weak
gradient pulse is infinitely long and the intense pulse is applied
instantaneously the spin echo attenuation is given
by~\cite{Laun11PhysRevLett}
\begin{equation}
\label{eq:ln_final}
E(q\,|\,\delta_{\textup{L}}\rightarrow\infty,\delta_{\textup{N}}\rightarrow 0)=\overline{S_{\text{0}}(q)}\,,
\end{equation}
which yields the form factor $\overline{S_{\text{0}}(q)}$ averaged
over the pores in the sample.  Under experimental conditions the
requirements $\delta_{\textup{L}}\rightarrow\infty$ and
$\delta_{\textup{N}}\rightarrow 0$ cannot be full-filled. Fortunately,
the need for infinitely long gradient pulses can be relaxed to the
condition that the molecules traverse the pore space multiple times
such that\label{pp:inequality} $D\delta_{\textup{L}}\gg L^2$. However,
this less stringent requirement has been linked to blurring of the
obtained average pore
image~\cite{Laun12PhysRevE,Hertel15PhysRevE}. Secondly, the diffusion
encoding during $\delta_{\textup{N}}$ has to be accounted for as well,
since the intense gradient pulse cannot be applied instantaneously. On
average the molecules push away from the pore walls during
$\delta_{\textup{N}}$, which leads to edge enhancement similar to the
phenomenon in MRI~\cite{Laun12PhysRevE,Hertel15PhysRevE}. These two
effects together with experimental adaptations of MRPI may pose
challenges for the formulation of fitting functions especially when
the gradient pattern involves many pulses as shown in
fig.~\ref{fig:mrpi_vs_pgse} (b) and (c). Therefore, the Multiple
Correlation Function (MCF) technique was employed for the simulation
of the spin echo decay. This simulation technique is flexible enough
for most gradient pulse patterns and it can adequately account for
both blurring and edge enhancement~\cite{Grebenkov07RevModPhys}.

In MCF the PGSE NMR signal is found by evaluating the ordered matrix
product
\begin{equation}
\label{eq:mcf_final}
E\approx \left[\prod^{\text{K}}_{\text{k=1}}\exp\left[-(p\Lambda+\im q_{\text{G}} f(k\tau)\mathcal{B})\tau/T\right]\right]_{\text{0},\text{0}}\,,
\end{equation}
where $[\dots]_{0,0}$ denotes the first diagonal element of the
resulting matrix. The piece-wise constant function $f(k\tau)$ is the
dimensionless temporal profile of the normalized spatial magnetic
field profile $\widetilde{B}(\vek{r})$ with maximum amplitude $\beta$,
such that the magnetic field at any time $t$ is given by
\begin{equation}
  B(\vek{r},k\tau)=f(k\tau)\beta \widetilde{B}(\vek{r})\,.
\end{equation}
$\mathcal{B}$ is the correlation matrix of the spatial gradient
profile $\tilde{B}(\vek{r})$ with the eigenfunctions of the Laplace
operator and $\Lambda$ is the diagonal eigenvalue matrix of the
Laplace operator. Furthermore, $p=DT/L^2$ is the reduced
self-diffusion coefficient and $q_{\textup{G}}=\gamma\beta TL$ denotes
the generalized gradient intensity. More detailed descriptions of the
MCF technique and its application to MRPI can be found
in~\cite{Laun12PhysRevE, Hertel15PhysRevE}.

For calibration purposes the spin echo decay has been measured
previously in an isotropic liquid~\cite{Hertel15PhysRevE}. When MRPI
is applied to nuclear spins in an isotropic unrestricted liquid one
obtains a Gaussian attenuation similar to other PGSE NMR techniques
given by
\begin{equation}
  \label{eq:gaussian}
  E(q)=\exp\{-D(2\pi q)^2\varDelta_{\textup{eff}}\},
\end{equation}
where $D$ is the self-diffusion coefficient of the spin bearing
molecules and $\varDelta_{\textup{eff}}$ is the effective observation
time, which is a function of the pulsed gradient pattern $G(t)$. In
the NMR self-diffusion literature, the term $(2\pi
q)^2\varDelta_{\textup{eff}}$ is usually captured as one variable in
the so called
$b$-factor~\cite{Stallmach07AnnuRepNMRSpectrosc}. However, in this
contribution the case of unrestricted self-diffusion will be compared
to the case of $q$-space imaging with MRPI. Therefore, we chose to
express the spin echo attenuation as a function of the $q$-value. The
effective diffusion time $\varDelta_{\textup{eff}}$ can be calculated
in analogy to the well known double integral for the calculation of
the $b$-factor in the PGSE NMR
literature~\cite{Stejskal65JCP_I,Stallmach07AnnuRepNMRSpectrosc}. It
is given by
\begin{equation}
\label{eq:double_int}
\varDelta_{\textup{eff}}=\frac{1}{(2\delta^{'})^2}\int_0^t\mathrm{d}t^{'}\left[\int_0^{t^{'}}\mathrm{d}t^{''}f(t^{''})\right]^2\,,
\end{equation}
where the factor $1/(2\delta^{'})^2$ is appearing, because it has been
factored into the $q$-value which is given by $q=\gamma
2\delta^{'}G_{\textup{N}}/2\pi$ for our
implementation. \label{pp:delta_eff} For this work, only two intense
gradient pulses were applied and $N_{\textup{L}}$ gradient pulses are
adding up to yield the weak long gradient. \label{pp:comment_16} Evaluation 
of eq.~\ref{eq:double_int} for the CPMG based MRPI pulse
sequence as shown in fig.~\ref{fig:mrpi_vs_pgse} (c)
yields~\cite{Hertel15PhysRevE}
\begin{equation}
  \label{eq:delta_eff}
  \varDelta_{\textup{eff}}=\frac{1}{3}\left[\delta^{'}(N_{\text{L}}{+}2){+}\delta_{\text{s}}\left(\frac{2N^2_{\text{L}}{+}3N_{\text{L}}{+}1}{2N_{\text{L}}}{+}\frac{3}{4}\right)\right]\,,
\end{equation}
where $\delta^{'}$ is the gradient pulse duration and where
$\delta_{\text{s}}$ is the separation between the gradient pulses,
i.e. the time from the end of one gradient pulse until the leading
edge of the next gradient pulse.

\section{Experimental parameters and samples}
\label{sec:exp}

\subsection{Pulse sequence design}
\label{sec:design}

\begin{figure}[htb]
  \centering 
  \setlength{\unitlength}{0.07mm}
  
\begin{picture}(1200,450)(-110,-100)
  \put(-80,120){\makebox[0mm][c]{\textbf{a)}}}%
  \Achse{0,50}{1,0}{1100}{0,-5}{$t$}
  \RFPuls{0,50}{170}{0.5mm}{3,13}{\small{$90^{\circ}_{0}$}}
  \MassPfeile{0,0}{1,0}{100}{-20}{0,-50}{$\tau$}
  \RFPuls{100,50}{170}{1mm}{5,15}{\small{$180^{\circ}_{\phi_{\text{P}}}$}}
  \MassPfeile{100,0}{1,0}{200}{-20}{0,-55}{\small{$2\tau$}}
  \RFPuls{300,50}{170}{1mm}{5,15}{\small{$180^{\circ}_{\frac{\pi}{2}}$}}
  \MassPfeile{300,0}{1,0}{200}{-20}{0,-55}{\small{$2\tau$}}
  \RFPuls{500,50}{170}{1mm}{5,15}{\small{$180^{\circ}_{\phi_{\text{P}}}$}}
  \MassPfeile{500,0}{1,0}{200}{-20}{0,-55}{\small{$2\tau$}}
  \RFPuls{700,50}{170}{1mm}{5,15}{\small{$180^{\circ}_{\frac{\pi}{2}}$}}
  \MassPfeile{700,0}{1,0}{200}{-20}{0,-55}{\small{$2\tau$}}
  \RFPuls{900,50}{170}{1mm}{5,15}{\small{$180^{\circ}_{\phi_{\text{P}}}$}}
  \MassPfeile{900,0}{1,0}{100}{-20}{0,-50}{$\tau$}
  \color{grey}
  \Echo{1000,50}{120}{0,0}{}{0,0}{}
  \color{black}
  \Echo{1000,50}{90}{10,30}{$\phi_{\textup{rec}}$}{0,0}{}
  \color{grey}
  \Echo{1000,50}{50}{0,0}{}{0,0}{}
  \color{black}
  \put(1000,40){\line(0,1){20}}
  \put(0,0){\line(0,1){50}}
  \put(100,0){\line(0,1){50}}
  \put(300,0){\line(0,1){50}}
  \put(500,0){\line(0,1){50}}
  \put(700,0){\line(0,1){50}}
  \put(900,0){\line(0,1){50}}
  \put(1000,0){\line(0,1){50}}
\end{picture}
\begin{picture}(1200,350)(-110,-100)
  \Achse{0,60}{0,1}{180}{0,0}{$\vek{G}(t)$}
  \put(0,20){\makebox[0mm][c]{\small{$t{=}0$}}}

  \put(-80,130){\makebox[0mm][c]{\textbf{b)}}}%
  \Achse{0,100}{1,0}{1100}{0,-5}{$t$}
  \color{grey}
  \PFGGrad{150,100}{Neg}{100}{30}{0}{-60,20}{}
  \PFGGrad{350,100}{Pos}{100}{30}{0}{-60,-30}{}
  \color{black}

  \color{grey}
  \PFGGrad{150,100}{Neg}{100}{10}{0}{-60,20}{}
  \PFGGrad{350,100}{Pos}{100}{10}{0}{-60,-30}{}
  \color{black}

  \color{black}
  \PFGGrad{150,100}{Neg}{100}{20}{0}{-60,20}{}
  \PFGGrad{350,100}{Pos}{100}{20}{0}{-60,-30}{}
  \color{black}

  \MassPfeile{150,125}{1,0}{100}{-10}{0,10}{$\delta^{'}$}
  \PFGGrad{10,100}{Pos}{50}{10}{0}{-50,20}{}
  \PFGGrad{950,100}{Pos}{100}{10}{0}{-50,20}{}

  \color{grey}
  \PFGGrad{550,100}{Pos}{100}{150}{0}{-60,30}{}
  \PFGGrad{750,100}{Neg}{100}{150}{0}{-60,30}{}
  \color{black}

  \color{grey}
  \PFGGrad{550,100}{Pos}{100}{50}{0}{-60,30}{}
  \PFGGrad{750,100}{Neg}{100}{50}{0}{-60,30}{}
  \color{black}

  \color{black}
  \PFGGrad{550,100}{Pos}{100}{100}{0}{-60,30}{}
  \PFGGrad{750,100}{Neg}{100}{100}{0}{-60,30}{}
  \color{black}

  \MassPfeile{750,125}{1,0}{100}{-10}{0,10}{$\delta^{'}$}

  \linethickness{.5mm}
  \put(93,40){\dashbox{1}(400,500)[]{}}
  \linethickness{.07mm}
  \put(250,-5){\makebox(30,-10)[l]{loop $\frac{N_{\text{L}}}{2}$}}
  \MassPfeile{855,-49}{0,1}{150}{10}{85,0}{$G_{\text{N}}$}
  \put(985,150){\vector(0,-1){40}}
  \put(985,60){\vector(0,1){40}}
 \put(1005,125){$G_{\text{R}}$}
  \put(360,170){\vector(0,-1){40}}
  \put(360,60){\vector(0,1){40}}
\put(375,150){$G_{\text{L}}$}
\end{picture}
\caption{MRPI pulse sequence with a CPMG like rf-pulse scheme (a). The
  long gradient is replaced by $N_{\text{L}}$ gradient pieces with
  gradient amplitude $G_{\text{L}}$ and the narrow gradient pulse was
  split into two intense gradient pulses with gradient amplitude
  $G_{\text{N}}$. The gradients are shown relative to each other as
  applied in the laboratory reference frame ($\vek{G}(t)$)
  (b). Effectively, the gradient pulses $G_{\text{L}}$ add up and are
  balanced by the two gradient pulses $G_{\text{N}}$. In practice the
  gradient pulses are ramped in order to reduce eddy currents (not
  shown).}
  \label{fig:mrpi_seq}
\end{figure}
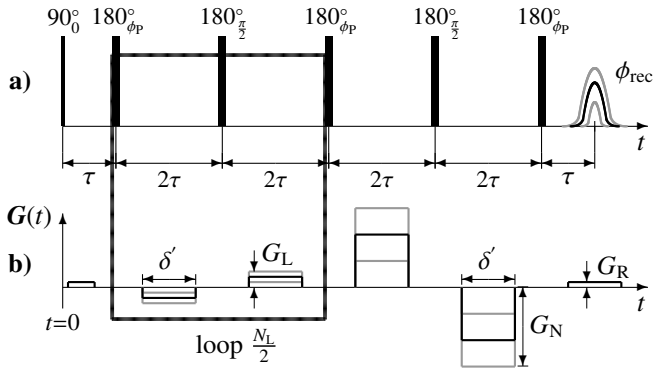

Figure \ref{fig:mrpi_seq} a) shows the rf-pulse sequence of the CPMG
based MRPI experiment~\cite{Hertel13PhysRevE,Hertel15PhysRevE}.
The rf-pulse phases were implemented according to the PIETA phase
scheme, where the odd $180\degree$ rf-pulses are phase cycled with
$\phi_{\text{P}}$. The phase in units of radians of the $90\degree$
rf-pulse was fixed at $0$, the phases of the even numbered
$180\degree$ rf-pulse were set to $\pi/2$ and the receiver phase
$\phi_{\textup{rec}}$ was kept at $0$. The gradient pulse scheme
$G(t)$ is depicted in fig.~\ref{fig:mrpi_seq} b). Here, $G(t)$ shows
the relative amplitude of the gradient pulses as they are applied in
the laboratory reference frame. There is a read gradient
$G_{\textup{R}}$ applied in the beginning of the pulse sequence and
during the detection period for the detection of gradient pulse
mismatches. For this study, the number of long gradient pulses with
amplitude $G_{\textup{L}}$ ranged from $N_{\textup{L}}=30$ to
$N_{\textup{L}}=50$. Note that no spin echo is formed until after the
last $180\degree$ rf-pulse. Thus, the spin echo count is fixed and for
the implementation shown in fig.~\ref{fig:mrpi_seq} it is given by
$n_{\text{E}}=N_{\textup{L}}+3$.\label{pp:eq_delta_p} Considering
eq.~\ref{eq:direct_selection} and the fact that $n_{\textup{E}}$ is an
odd number it follows that the desired signal is contained in the
matrix elements of $s(q,t,\Delta P)$ where $\Delta
P=-(n_{\textup{E}}+1)=-(N_{\textup{L}}+4)$.

Previous MRPI results as discussed in \cite{Hertel13PhysRevE} and
~\cite{Hertel15PhysRevE} have been obtained with the same pulse
sequence, but with a $4$-step phase cycle. Table~\ref{tab:phase_cycle}
shows the phase cycle, which has been designed in analogy to a cyclops
phase cycle adapted to the special case of the MRPI pulse sequence.
\begin{table}[htb]
  \centering
  \begin{tabular}{ccccc}
    \hline\noalign{\smallskip}
    \multicolumn{4}{c}{Pulse phases} & \\
    \cline{1-4}\noalign{\smallskip}
    $\phi_{90}$ & $\phi_{\textup{odd}}$ & $\phi_{\textup{even}}$ & $\phi_{\textup{last}}$ & $\phi_{\textup{rec}}$\\
    \hline\\
    $0$ & $0$ & $\pi$ & $\frac{\pi}{2}$ & $\pi$\\
    $0$ & $\frac{\pi}{2}$ & $\frac{3\pi}{2}$ & $0$ & $0$\\
    $0$ & $\pi$ & $0$ & $\frac{3\pi}{2}$ & $\pi$\\
    $0$ & $\frac{3\pi}{2}$ & $\frac{\pi}{2}$ & $\pi$ & $0$\\
    \hline
   \end{tabular}
   \caption{Phase cycle of the CPMG based MRPI pulse sequence with $4$ phase steps. The rf-phases $\phi_{\text{odd}}$ correspond to the odd numbered $180\degree$ rf-pulses 
     in fig.~\ref{fig:mrpi_seq} (a), while $\phi_{\text{even}}$ is the phase applied for the even numbered $180\degree$ rf-pulses. The last $180\degree$ pulse has a separate phase $\phi_{\textup{last}}$ assigned.}
  \label{tab:phase_cycle}
\end{table}
The rf-pulse phases are not shown in fig.~\ref{fig:mrpi_seq}, but have
been named $\phi_{\textup{odd}}$ for the odd $180\degree$ rf-pulses,
$\phi_{\textup{even}}$ for the even $180\degree$ rf-pulses and
$\phi_{\textup{last}}$ for the last $180\degree$ rf-pulse before
detection. The rf-pulse phase $\phi_{\textup{last}}$ has been
introduced to filter the pathway $p=0,+1,0,\dots 0,-1$, which does not
experience the stepped gradient pulses and can therefore not be
dephased.

\subsection{Hardware and samples}
\label{sec:hardware}

The experiments were carried out on a Bruker Avance II spectrometer
with a proton resonance frequency of $400\,\mega\hertz$. The gradient
pulses were generated by a Bruker GREAT 60 gradient amplifier in
conjunction with a Micro 2.5 imaging probe. At maximum current output
of $60\,\ampere$ the gradient system provided gradient strengths up to
$G_{\textup{max}}=\pm 1.45\,\tesla\metre^{-1}$. The utilized rf-coil
had a birdcage coil design with an inner diameter of
$ID=1\,\centi\metre$. Typical rf-pulse durations were
$t_{\textup{90\degree}}=12.5\,\micro\second$ and
$t_{\textup{180\degree}}=25\,\micro\second$ when utilizing
$^1\text{H}$ as the NMR active nuclei.

A spherical sample tube ($OD=1\,\centi\metre$) filled with
$\text{CuSO}_{\textup{4}}$ doped $\text{H}_{\textup{2}}\text{O}$ was
utilized for testing of the MRPI pulse sequence and to measure the
spin echo attenuation for unrestricted self-diffusion. For the MRPI
experiments in a model porous system a capillary sample was prepared,
which contained ca. $470$ polyimide coated glass capillaries with
inner diameter of $d=20\pm 2\,\micro\metre$. These capillaries were
filled with distilled $\text{H}_{\textup{2}}\text{O}$ according to the
procedure outlined in~\cite{Hertel15PhysRevE}.

\section{Results}
\label{sec:results}

\begin{figure}[htb]
  \centering
  \includegraphics[width=.47\textwidth]{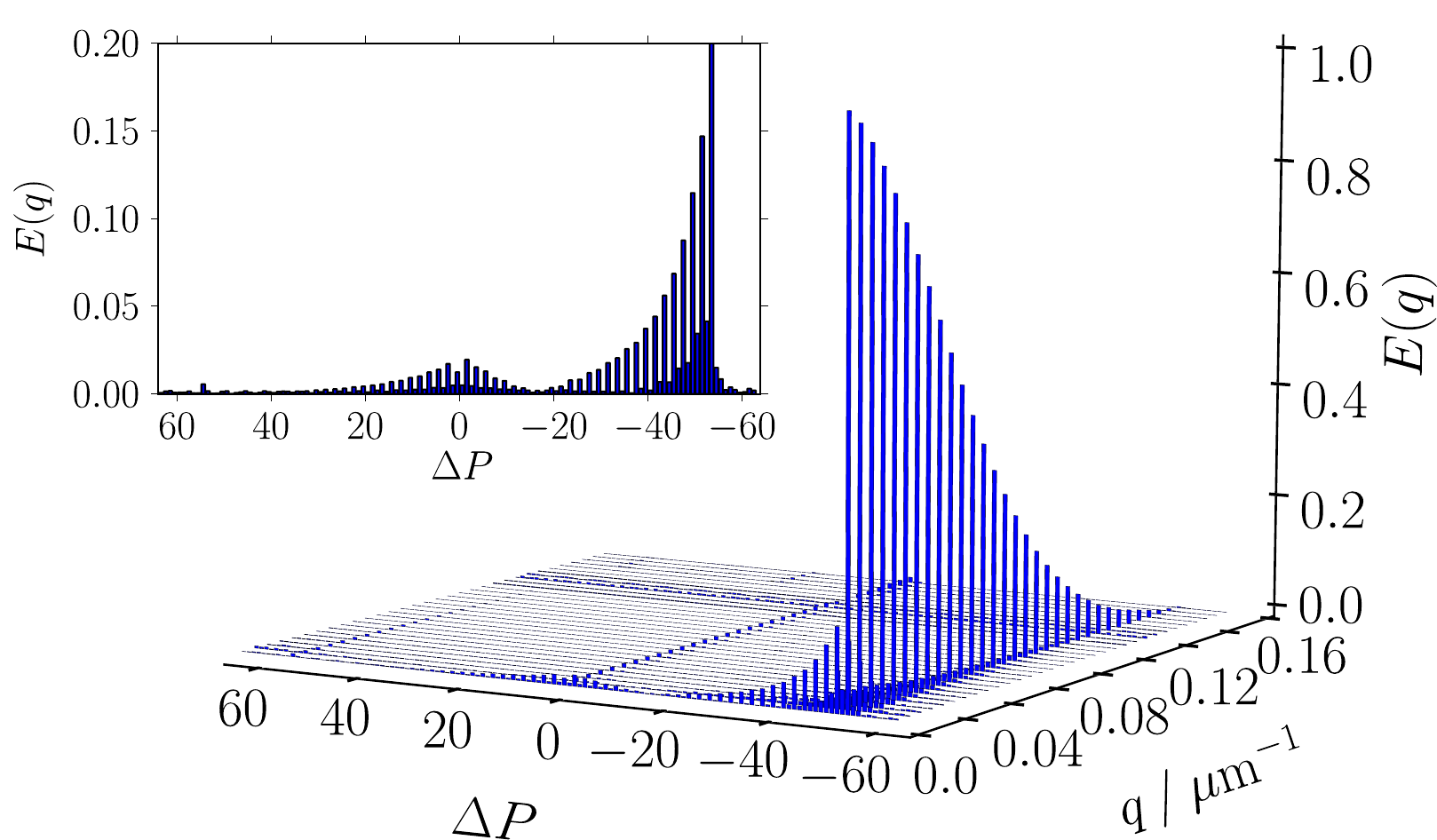}
  \caption{Coherence transfer pathway map obtained by applying MRPI
    including the PIETA phase cycling scheme to the spherical water
    sample. Shown is the real part of signal matrix at the center of
    the spin echo as a function of the cumulative coherence change
    $\Delta P$. The desired pathway is the direct coherence pathway
    with $\Delta P=-54$. The inset shows the CP spectrum for zero
    pulsed field gradient amplitude ($q=0$).}
  \label{fig:3dcp_freediff}
\end{figure}

It has been shown previously that the MRPI pulse sequence can be
calibrated by measuring the free self-diffusion coefficient $D$ of a
suitable isotropic liquid~\cite{Hertel15PhysRevE}. Thus, the CPMG
based MRPI pulse sequence was applied to the spherical
$\text{H}_{\textup{2}}\text{O}$ sample described in
sec.~\ref{sec:hardware} and the PIETA phase cycle was applied to
investigate the coherence pathway spectrum. The number of long
gradient pulses was $N_{\textup{L}}=50$, which were refocused by
$N_{\textup{N}}=2$ narrow gradient pulses. The measured spin echo
corresponds to the spin echo count $n_{\textup{E}}=53$ and therefore
the number of phase steps was set to $N=128$, respecting the condition
of eq.~\ref{eq:Nscan_pieta}. The resulting data matrix
$s(q,t,\phi_{\textup{P}})$ had dimensions of $32\times 512\times 128$,
where $\bm{q}$ was incremented in $32$ steps, $512$ points were
acquired in the time domain for each scan and $128$ phase steps were
acquired. A subsequent Fourier transform analogous to eq.~\ref{eq:FT}
in the $\phi_{\textup{P}}$ direction yielded the signal matrix
$s(q,t,\Delta P)$. Fig.~\ref{fig:3dcp_freediff} shows the signal
matrix with dimension $32\times 128$, where the instance in time $t$
is chosen to show only the spin echo centers.
The coherence pathway spectrum shows several peaks in the $\Delta P$
dimension for the first row in the spectrum where $q=0$. Since the
contribution of unwanted pathways is the highest for zero gradient
amplitude the corresponding $\Delta P$ spectrum is shown separately in
the inset. The direct coherence pathway can be selected at $\Delta
P=-54$ according to eq.~\ref{eq:direct_selection}. It is apparent that
the attenuation of the pathway with $\Delta P=-54$ follows a Gaussian
shape as expected by eq.~\ref{eq:gaussian}, while all other pathways
experience a sharper drop in amplitude when $q$ is increased. This is
a consequence of their dephasing with increasing gradient pulse
amplitudes. One exception being the CP at $\Delta P=0$, with
$\tilde{\vek{p}}=(0,+1,0,\dots,0,-1)$. This CP is not dephased, since
it is only influenced by the constant read gradient in the beginning
and the end of the pulse sequence $G_{\textup{R}}$ for which the spin
echo condition $\int_{0}^{T}G_{\textup{R}}^*(t)=0$ is full-filled
regardless of the amplitude of the other pulsed gradients. This
pathway was accounted for when designing the $4$-step phase cycling
scheme discussed in sec.~\ref{sec:design}. Furthermore, the relative
amplitude of all signal contributions at low $q$ values is high enough
to significantly impact on the spin echo amplitude if they are not
filtered in an experiment without PIETA. The most important pathways
are the ones close to $\Delta P=0$, which are stimulated echo pathways
with many intervals where the magnetization is stored in the
$z$-direction (i.e. many elements with $p_{\text{k}}=0$), while
coherence pathways close to the direct pathway with $\Delta P=-54$
also appear with significant intensity. The latter pathways are
characterized by a small number of time intervals where the
magnetization is stored in $z$-direction. During the $z$-storage time
intervals the magnetization associated with these CPs is subject to
$T_{\textup{1}}$ relaxation instead of $T_{\textup{2}}$
relaxation. Thus, their relative amplitude with respect to the direct
CP may be very high even in the case of nearly perfect rotations of
the magnetization.

\begin{figure}[htb]
  \centering
  \includegraphics[width=.47\textwidth]{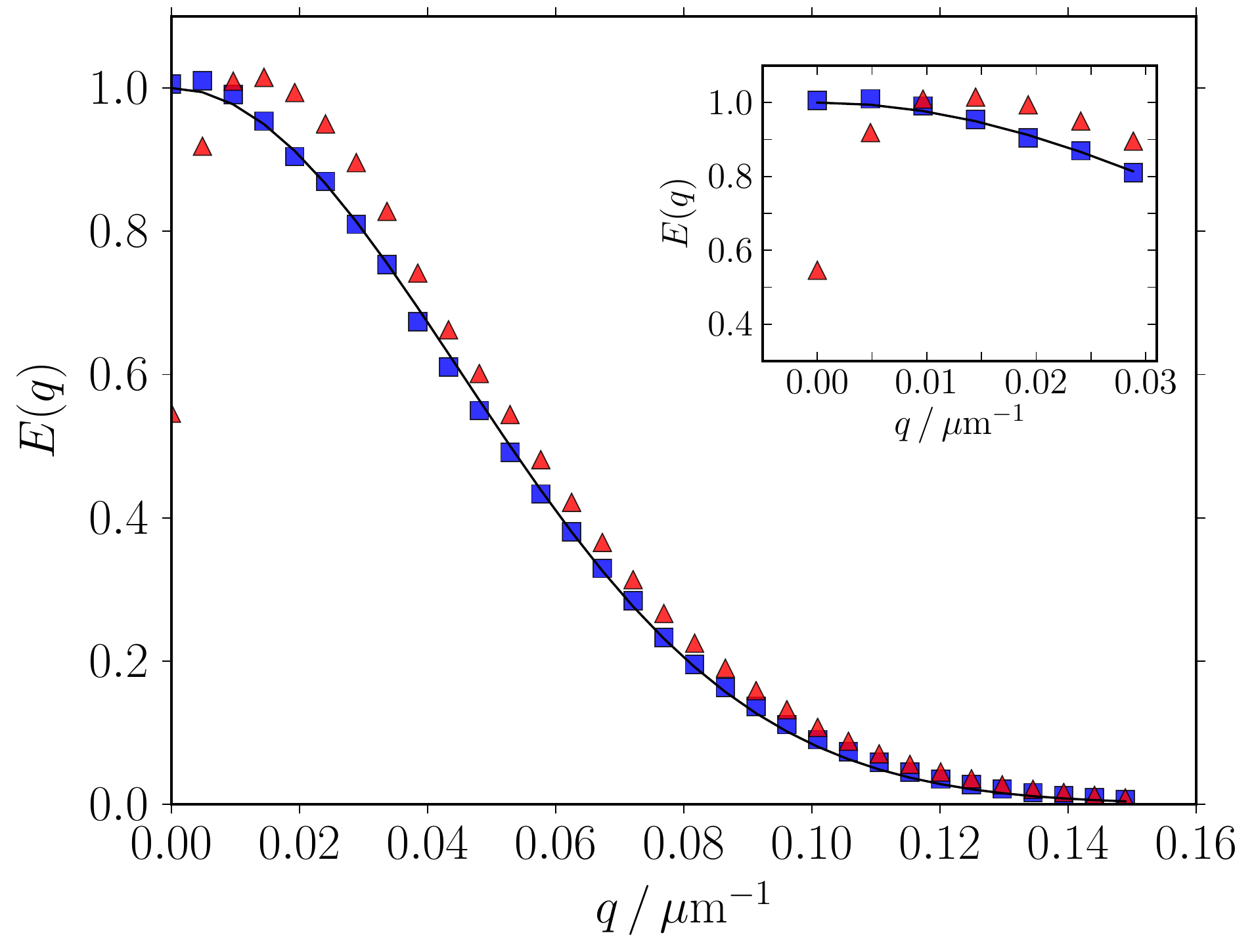}
  \caption{Comparison of the MRPI spin echo attenuation between the
    $4$-step phase cycling scheme (triangles) and the PIETA phase
    cycling scheme with subsequent selection of the direct coherence
    pathway (squares). The insert shows a zoom of the signal
    amplitudes at low $q$ values.}
  \label{fig:pieta_vs_4step}
\end{figure}

Fig.~\ref{fig:pieta_vs_4step} shows the normalized spin-echo attenuation
$E(q)$ as obtained by selecting the direct coherence pathway with
$\Delta P=-54$ (squares) in comparison to an MRPI experiment performed
with the $4$-step phase cycle (triangles).
All parameters were kept the same for both experiments except for the
phase cycling scheme. For both experiments the gradient duration was
$\delta^{'}=2\,\milli\second$ and the inter $180\degree$ rf-pulse spacing
was $2\tau=2.5\,\milli\second$.  The solid line in
fig.~\ref{fig:pieta_vs_4step} represents the fit of the experimental
data against eq.~\ref{eq:gaussian}. The fitted self-diffusion
coefficient was \mbox{$D=(2.8\pm 0.2)\times
  10^{-9}\metre^2\second^{-1}$}, which is higher than the literature
value for $T=293\,\kelvin$ of \mbox{$D=(2.0\pm 0.2)\times
  10^{-9}\metre^2\second^{-1}$}~\cite{Holz00PCCP}. However, it has
been shown that the right self-diffusion coefficient can be obtained
with MRPI when less gradient pulses are
utilized~\cite{Hertel15PhysRevE}. This suggests that the gradient ramp
times which are not accounted for in eq.~\ref{eq:delta_eff} may play
an increasingly important role with increasing numbers of gradient
pulses. This is consistent with findings for pulse sequences
specifically designed for the measurement of self-diffusion
coefficients~\cite{Price91JMR}.

Nonetheless, the expected Gaussian attenuation is obtained and the
effect of the coherence pathways on the spin echo amplitude for low
$q$ values can be studied. Clearly, the amplitudes of the first four
data points are lower than expected for the $4$-step phase cycle
(triangles). Especially the fist point measured with zero gradient
amplitude is $45\%$ below the expected amplitude. This signal
suppression is caused by spurious coherence pathways which interfere
with the direct coherence pathway destructively. It may appear
surprising that the spin echo amplitude first increases before it
asymptotically follows the expected Gaussian attenuation according to
eq.~\ref{eq:gaussian}. This surprising increase in amplitude is caused
by the fact that the recorded amplitude is obtained by the sum of the
individual CPs. Each CP is associated with a corresponding complex
magnetization. When the magnetization of the CPs is recorded with a
different magnetization phase during detection, they do interfere
destructively. On the other hand, the spin echo attenuation extracted
from the MRPI PIETA experiment does not suffer from destructive
interference and thus the correct amplitude for low $q$-values was
recovered.

Following the bulk experiments, MRPI PIETA was applied to the
capillary sample with the gradient pulses applied perpendicular to the
cylinder axis. The number of long gradient pulses was
$N_{\textup{L}}=30$ and therefore $n_{\textup{E}}=33$. 

\begin{figure}[htb]
  \centering
  \includegraphics[width=.47\textwidth]{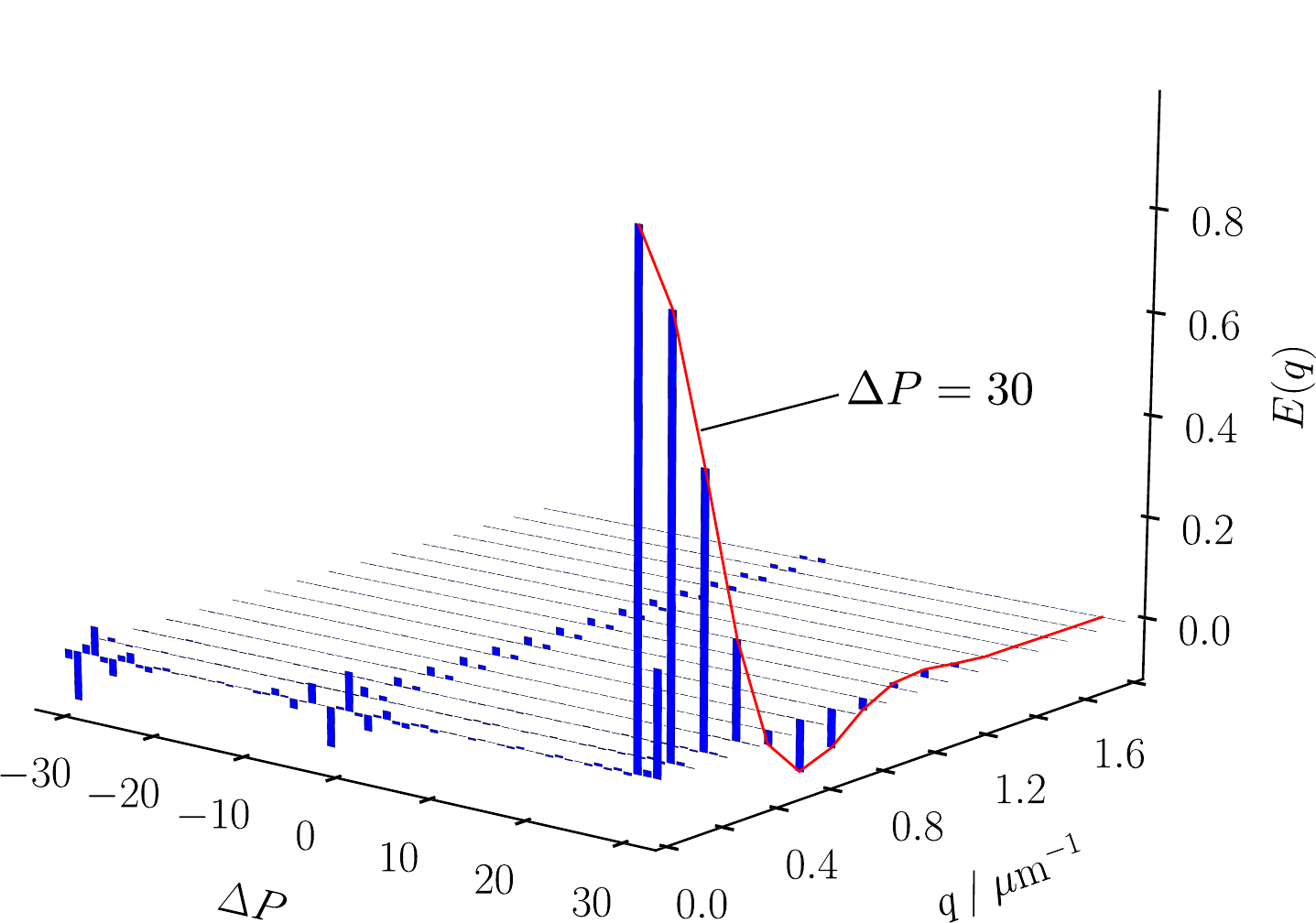}
  \caption{Coherence transfer pathway map obtained by applying MRPI
    including the PIETA phase cycling scheme to the cylindrical
    capillary sample. Real part of the coherence pathways with total
    coherence change $\Delta P$ for the center of the spin echo. The
    desired pathway is the direct coherence pathway with $\Delta
    P=-34$ which appears at $\Delta P=30$ due to fold-back.}
  \label{fig:3dcp_cyl}
\end{figure}

The number of phase steps was set to $N=64$, which was $2$ phase steps
less than required by eq.~\ref{eq:Nscan_pieta}. This was due to the
requirement of FFT algorithms to operate on sampling points of
multiples of power of two, else the necessary number of phase steps
would have been $N=128$. This would have extended the experimental
time beyond practical limits. In hindsight, a Bluestein
algorithm~\cite{Bluestein70IEEE} accepting an arbitrary number of
samples would be favorable in the future. Analysis of the results,
however, can still be carried out with the under sampled data set,
since fold-back signals do not overlap in the resulting coherence
pathway spectrum. Fig.~\ref{fig:3dcp_cyl} shows the real part of the
matrix $s(q,\Delta P)$ with the dimension $16\times 64$, which was
obtained from the matrix $s(q,t,\Delta P)$ by selecting the signal at
the time instance of the spin echo center.
Further parameters were the gradient duration of
$\delta^{'}=4.635\,\milli\second$ and the echo time of
$2\tau=5.56\,\milli\second$. Due to the under sampling of the spectrum
in the $\Delta P$ dimension, one obtains a CP spectrum with fold backs
of the desired signal $\Delta P=-34$ which appears at $\Delta P =
30$. Similar to the experiment shown for the free self-diffusion case
there are several coherence pathways close to $\Delta P = -32$ and in
the centre of the spectrum at $\Delta P=0$ for
$q=0$. Here, however also CPs close to $\Delta P=+32$
are contributing significantly. These pathways are initially following
the pathway $\vek{p}=(0,-1,+1,\dots,-1)$ which is conjugate to the
desired pathway $\tilde{\vek{p}}=(0,+1,-1,\dots,-1)$ and they are
transferred with the last few $180\degree$ rf-pulses into coherence
pathways which are detectable. These CPs appear in the cylindrical
capillary spectrum, while they do not contribute appreciably to the
bulk $\text{H}_{\textup{2}}\text{O}$ spectrum shown in
fig.~\ref{fig:3dcp_freediff}.  This may be explained by higher
internal gradients in the capillaries which render the rf-pulse
rotations less ideal than in the bulk sample. Thus, the contributions
of coherence pathways depends on the sample as well as the careful
adjustment of the rf-pulse durations. Regardless, their relative
importance can be measured directly with PIETA and the desired
coherence pathway can be extracted.

\begin{figure}[htb]
  \centering
  \includegraphics[width=.47\textwidth, clip]{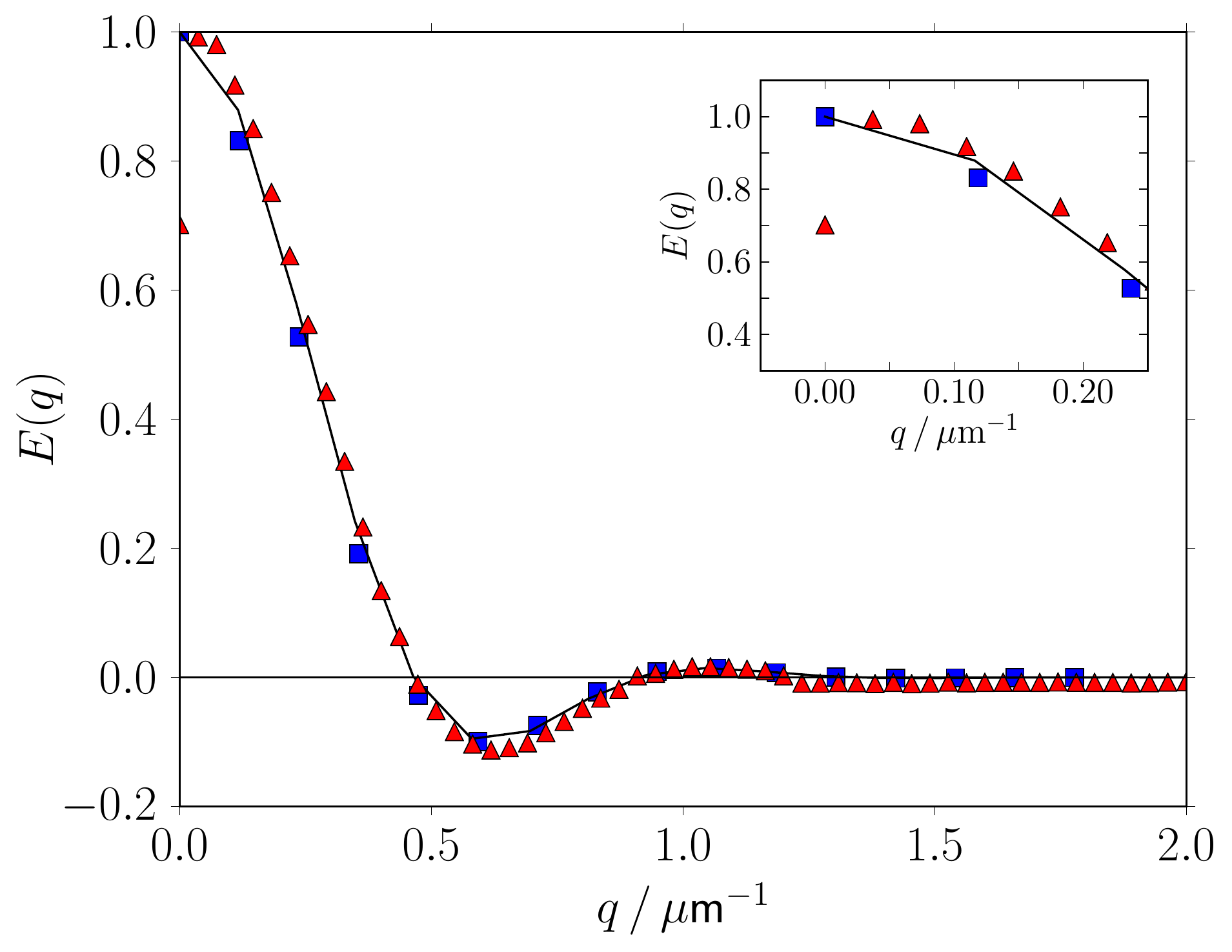}
  \caption{Real part of the $\bm{q}$-space profile of the capillary
    sample as obtained by the MRPI pulse sequence. The spin echo
    amplitude for the experiment utilizing the $4$-step phase cycle
    (triangles) is distorted at low gradient strengths due to
    destructive coherence pathways. The solid line was obtained by MCF
    simulation~\cite{Hertel15PhysRevE}. The insert shows a zoom of the
    signal amplitudes at low $q$ values.}
  \label{fig:pieta_vs_4step_cylinder}
\end{figure}

Figure~\ref{fig:pieta_vs_4step_cylinder} (squares) shows the real part
of the signal as obtained by selecting the direct coherence pathway
appearing at $\Delta P=30$.
The spin echo amplitude for the $4$-step phase cycle (triangles) is
suppressed by $30\%$ below the expected amplitude at zero gradient
strength ($q=0$). The $q$-values in this experiment are over a factor
of $10$ larger than in the experiments with the bulk
$\text{H}_{\textup{2}}\text{O}$ sample shown in
fig.~\ref{fig:pieta_vs_4step}. Therefore, only the first and second
data points are affected by the destructive coherence pathways, while
for all the following $q$-steps these CPs have been
dephased. Furthermore, the number of $180\degree$ rf-pulses is only
$n_{\textup{180}}=33$ versus $n_{\textup{180}}=53$ in the bulk
experiments shown in fig.~\ref{fig:pieta_vs_4step}. Thus, the relative
importance of undesired coherence pathways to the desired coherence
pathway is higher in experiments with more rf-pulses.

The solid line in fig.~\ref{fig:pieta_vs_4step_cylinder} is the
simulated signal as obtained with the Multiple Correlation Function
technique~\cite{Hertel15PhysRevE}. The signal extracted from the PIETA
experiment recovers the signal amplitudes for low values of
$q$. Furthermore, the signal phase is preserved with this phase
cycling technique as can be seen by the positive and negative lobes in
fig.~\ref{fig:pieta_vs_4step_cylinder}. Thus, PIETA may not only prove
useful for relaxation studies in CPMG like experiments, but it can
also help with the development of imaging techniques where phase
preservation through a sequence of $180\degree$ rf-pulses is a major
concern. Such techniques are for example Rapid Imaging with Refocused
Echoes (RARE) experiments~\cite{Hennig86MagnResonMed} and spatially resolved
$T_{\text{2}}$ experiments~\cite{Li09JMR_II}.

\section{Conclusions}

It has been shown that PIETA can recover the desired NMR signal in
cases where a broad coherence pathway spectrum complicates the design
of complete phase cycles with the nested phase cycling approach. The
results with MRPI PIETA show that the NMR signal phase due to gradient
pulses can be preserved with this phase cycling technique, which may
prove beneficial for applications involving other NMR imaging
techniques. The only two steps required for implementing PIETA are to
cycle every second rf-pulse phase and to respect the Nyquist-Shannon
criterion for the number of phase increments recorded. This
standardized scheme of implementation may shorten the time required
for pulse sequence design by allowing one to extract the desired
signal without a deep analysis of the CP spectrum. However, PIETA
requires an extra Fourier transform processing step as compared to
Cogwheel phase cycling. Thus, Cogwheel phase cycling may be more
efficient during routine application of pulse sequences where the
extra information about contributing coherence pathways is not
needed. Prime candidates for applications of PIETA are CPMG like pulse
sequences. Such pulse sequence designs are increasingly utilized for a
broad range of NMR experiments. Examples involving spin-$1/2$ nuclei
include the spectral characterization of the self-diffusion of
molecules~\cite{Callaghan95JMRComm}, compartment sizing in
MRI~\cite{Shemesh13JMR} and the enhancement of the signal in low field
NMR. It can be assumed that the PIETA approach will prove beneficial
for these applications in the future.

\section*{Acknowledgements}

We gratefully acknowledge financial support by the Ministry of
Business, Innovation and Employment of New Zealand. S.H. thanks
Victoria University of Wellington for receiving support through a
Ph.D. scholarship.
\\
\bibliographystyle{model1-num-names}

\begin{thebibliography}{38}
\expandafter\ifx\csname natexlab\endcsname\relax\def\natexlab#1{#1}\fi
\providecommand{\bibinfo}[2]{#2}
\ifx\xfnm\relax \def\xfnm[#1]{\unskip,\space#1}\fi
\bibitem[{Stejskal and Tanner(1965)}]{Stejskal65JCP_I}
\bibinfo{author}{E.~O. Stejskal}, \bibinfo{author}{J.~E. Tanner},
\newblock \bibinfo{title}{Spin diffusion measurements: Spin echoes in the
  presence of a time-dependent field gradient},
\newblock \bibinfo{journal}{J. Chem. Phys.} \bibinfo{volume}{42}
  (\bibinfo{year}{1965}) \bibinfo{pages}{288--292}.
\bibitem[{Callaghan(2011)}]{Callaghan11Book}
\bibinfo{author}{P.~T. Callaghan}, \bibinfo{title}{Translational Dynamics \&
  Magnetic Resonance}, \bibinfo{publisher}{Oxford University Press},
  \bibinfo{address}{Oxford}, \bibinfo{year}{2011}.
\bibitem[{Callaghan et~al.(1991)Callaghan, Coy, Macgowan, Packer, and
  Zelaya}]{Callaghan91Nature}
\bibinfo{author}{P.~T. Callaghan}, \bibinfo{author}{A.~Coy},
  \bibinfo{author}{D.~Macgowan}, \bibinfo{author}{K.~J. Packer},
  \bibinfo{author}{F.~O. Zelaya},
\newblock \bibinfo{title}{Diffraction-like effects in {NMR} diffusion studies
  of fluids in porous solids},
\newblock \bibinfo{journal}{Nature} \bibinfo{volume}{351}
  (\bibinfo{year}{1991}) \bibinfo{pages}{467}.
\bibitem[{Laun et~al.(2011)Laun, Kuder, Semmler, and
  Stieltjes}]{Laun11PhysRevLett}
\bibinfo{author}{F.~B. Laun}, \bibinfo{author}{T.~A. Kuder},
  \bibinfo{author}{W.~Semmler}, \bibinfo{author}{B.~Stieltjes},
\newblock \bibinfo{title}{Determination of the defining boundary in nuclear
  magnetic resonance diffusion experiments},
\newblock \bibinfo{journal}{Phys. Rev. Lett.} \bibinfo{volume}{107}
  (\bibinfo{year}{2011}) \bibinfo{pages}{048102}.
\bibitem[{Laun et~al.(2012)Laun, Kuder, Wetscherek, Stieltjes, and
  Semmler}]{Laun12PhysRevE}
\bibinfo{author}{F.~B. Laun}, \bibinfo{author}{T.~A. Kuder},
  \bibinfo{author}{A.~Wetscherek}, \bibinfo{author}{B.~Stieltjes},
  \bibinfo{author}{W.~Semmler},
\newblock \bibinfo{title}{{NMR}-based diffusion pore imaging},
\newblock \bibinfo{journal}{Phys. Rev. E} \bibinfo{volume}{86}
  (\bibinfo{year}{2012}) \bibinfo{pages}{021906}.
\bibitem[{Kuder et~al.(2013)Kuder, Bachert, Windschuh, and
  Laun}]{Kuder13PhysRevLett}
\bibinfo{author}{T.~A. Kuder}, \bibinfo{author}{P.~Bachert},
  \bibinfo{author}{J.~Windschuh}, \bibinfo{author}{F.~B. Laun},
\newblock \bibinfo{title}{Diffusion pore imaging by hyperpolarized xenon-129
  nuclear magnetic resonance},
\newblock \bibinfo{journal}{Phys. Rev. Lett.} \bibinfo{volume}{111}
  (\bibinfo{year}{2013}) \bibinfo{pages}{028101}.
\bibitem[{Hertel et~al.(2013)Hertel, Hunter, and Galvosas}]{Hertel13PhysRevE}
\bibinfo{author}{S.~Hertel}, \bibinfo{author}{M.~Hunter},
  \bibinfo{author}{P.~Galvosas},
\newblock \bibinfo{title}{Magnetic resonance pore imaging, a tool for porous
  media research},
\newblock \bibinfo{journal}{Phys. Rev. E} \bibinfo{volume}{87}
  (\bibinfo{year}{2013}) \bibinfo{pages}{030802}.
\bibitem[{Hertel et~al.(2015)Hertel, Wang, Coard, Simpson, Hunter, and
  Galvosas}]{Hertel15PhysRevE}
\bibinfo{author}{S.~A. Hertel}, \bibinfo{author}{X.~Wang},
  \bibinfo{author}{P.~Coard}, \bibinfo{author}{C.~Simpson},
  \bibinfo{author}{M.~Hunter}, \bibinfo{author}{P.~Galvosas},
\newblock \bibinfo{title}{{M}agnetic {R}esonance {P}ore {I}maging of
  microscopic non-symmetric pore shapes},
\newblock \bibinfo{journal}{Phys. Rev. E} \bibinfo{volume}{92}
  (\bibinfo{year}{2015}) \bibinfo{pages}{012808}.
\bibitem[{Goelman and Prammer(1995)}]{Goelman95JMR}
\bibinfo{author}{G.~Goelman}, \bibinfo{author}{M.~Prammer},
\newblock \bibinfo{title}{The {CPMG} pulse sequence in strong magnetic field
  gradients with applications to oil-well logging},
\newblock \bibinfo{journal}{J. Magn. Reson. A} \bibinfo{volume}{113}
  (\bibinfo{year}{1995}) \bibinfo{pages}{11 -- 18}.
\bibitem[{H\"urlimann(2001)}]{Huerlimann01JMR}
\bibinfo{author}{M.~D. H\"urlimann},
\newblock \bibinfo{title}{Diffusion and relaxation effects in general stray
  field {NMR} experiments},
\newblock \bibinfo{journal}{J. Magn. Reson.} \bibinfo{volume}{148}
  (\bibinfo{year}{2001}) \bibinfo{pages}{367--378}.
\bibitem[{Song(2002)}]{Song02JMR}
\bibinfo{author}{Y.~Q. Song},
\newblock \bibinfo{title}{Categories of coherence pathways for the {CPMG}
  sequence},
\newblock \bibinfo{journal}{J. Magn. Reson.} \bibinfo{volume}{157}
  (\bibinfo{year}{2002}) \bibinfo{pages}{82--91}.
\bibitem[{Baltisberger et~al.(2012)Baltisberger, Walder, Keeler, Kaseman,
  Sanders, and Grandinetti}]{Baltisberger12JChemPhys}
\bibinfo{author}{J.~H. Baltisberger}, \bibinfo{author}{B.~J. Walder},
  \bibinfo{author}{E.~G. Keeler}, \bibinfo{author}{D.~C. Kaseman},
  \bibinfo{author}{K.~J. Sanders}, \bibinfo{author}{P.~J. Grandinetti},
\newblock \bibinfo{title}{Phase incremented echo train acquisition in {NMR}
  spectroscopy},
\newblock \bibinfo{journal}{J. Chem. Phys.} \bibinfo{volume}{136}
  (\bibinfo{year}{2012}) \bibinfo{pages}{211104}.
\bibitem[{Levitt(2001)}]{Levitt01Book}
\bibinfo{author}{M.~H. Levitt}, \bibinfo{title}{Spin dynamics: {B}asics of
  {N}uclear {M}agnetic {R}esonance}, \bibinfo{publisher}{Wiley},
  \bibinfo{year}{2001}.
\bibitem[{Bain(1984)}]{Bain84JMR}
\bibinfo{author}{A.~D. Bain},
\newblock \bibinfo{title}{Coherence levels and coherence pathways in {NMR}. {A}
  simple way to design phase cycling procedures},
\newblock \bibinfo{journal}{J. Magn. Reson. 1969} \bibinfo{volume}{56}
  (\bibinfo{year}{1984}) \bibinfo{pages}{418 -- 427}.
\bibitem[{Bodenhausen et~al.(1984)Bodenhausen, Kogler, and
  Ernst}]{Bodenhausen84JMR}
\bibinfo{author}{G.~Bodenhausen}, \bibinfo{author}{H.~Kogler},
  \bibinfo{author}{R.~Ernst},
\newblock \bibinfo{title}{Selection of coherence-transfer pathways in {NMR}
  pulse experiments},
\newblock \bibinfo{journal}{J. Magn. Reson. 1969} \bibinfo{volume}{58}
  (\bibinfo{year}{1984}) \bibinfo{pages}{370 -- 388}.
\bibitem[{Hughes et~al.(2004)Hughes, Carravetta, and Levitt}]{Hughes04JMR}
\bibinfo{author}{C.~E. Hughes}, \bibinfo{author}{M.~Carravetta},
  \bibinfo{author}{M.~H. Levitt},
\newblock \bibinfo{title}{Some conjectures for cogwheel phase cycling},
\newblock \bibinfo{journal}{J. Magn. Reson.} \bibinfo{volume}{167}
  (\bibinfo{year}{2004}) \bibinfo{pages}{259 -- 265}.
\bibitem[{Jerschow and M\"uller(1998)}]{Jerschow98JMR}
\bibinfo{author}{A.~Jerschow}, \bibinfo{author}{N.~M\"uller},
\newblock \bibinfo{title}{Efficient simulation of coherence transfer pathway
  selection by phase cycling and pulsed field gradients in {NMR}},
\newblock \bibinfo{journal}{J. Magn. Reson.} \bibinfo{volume}{134}
  (\bibinfo{year}{1998}) \bibinfo{pages}{17 -- 29}.
\bibitem[{Ollerenshaw and McClung(2000)}]{Ollerenshaw00JMR}
\bibinfo{author}{J.~Ollerenshaw}, \bibinfo{author}{R.~McClung},
\newblock \bibinfo{title}{Construction of phase cycles of minimum cycle length:
  Make{C}ycle},
\newblock \bibinfo{journal}{J. Magn. Reson.} \bibinfo{volume}{143}
  (\bibinfo{year}{2000}) \bibinfo{pages}{255 -- 265}.
\bibitem[{McClung(1999)}]{McClung99ConMagRes}
\bibinfo{author}{R.~E.~D. McClung},
\newblock \bibinfo{title}{Coherence transfer pathways and phase cycles: The
  decoding of a pulse sequence},
\newblock \bibinfo{journal}{Conc. Magn. Reson.} \bibinfo{volume}{11}
  (\bibinfo{year}{1999}) \bibinfo{pages}{1--28}.
\bibitem[{Levitt and Freeman(1979)}]{Levitt79JMR}
\bibinfo{author}{M.~H. Levitt}, \bibinfo{author}{R.~Freeman},
\newblock \bibinfo{title}{{NMR} population inversion using a composite pulse},
\newblock \bibinfo{journal}{J. Magn. Reson. (1969)} \bibinfo{volume}{33}
  (\bibinfo{year}{1979}) \bibinfo{pages}{473 -- 476}.
\bibitem[{Bax et~al.(1980)Bax, De~Jong, Mehlkopf, and
  Smidt}]{Bax80ChemPhysLett}
\bibinfo{author}{A.~Bax}, \bibinfo{author}{P.~De~Jong},
  \bibinfo{author}{A.~Mehlkopf}, \bibinfo{author}{J.~Smidt},
\newblock \bibinfo{title}{Separation of the different orders of {NMR}
  multiple-quantum transitions by the use of pulsed field gradients},
\newblock \bibinfo{journal}{Chem. Phys. Lett.} \bibinfo{volume}{69}
  (\bibinfo{year}{1980}) \bibinfo{pages}{567--570}.
\bibitem[{Carr and Purcell(1954)}]{Carr54PhysRev}
\bibinfo{author}{H.~Y. Carr}, \bibinfo{author}{E.~M. Purcell},
\newblock \bibinfo{title}{Effects of diffusion on free precession in nuclear
  magnetic resonance experiments},
\newblock \bibinfo{journal}{Phys. Rev.} \bibinfo{volume}{94}
  (\bibinfo{year}{1954}) \bibinfo{pages}{630--638}.
\bibitem[{Meiboom and Gill(1958)}]{Meiboom58RSI}
\bibinfo{author}{S.~Meiboom}, \bibinfo{author}{D.~Gill},
\newblock \bibinfo{title}{Modified spin-echo method for measuring nuclear
  relaxation times},
\newblock \bibinfo{journal}{Rev. Sci. Instrum.} \bibinfo{volume}{29}
  (\bibinfo{year}{1958}) \bibinfo{pages}{688--691}.
\bibitem[{Levitt et~al.(2002)Levitt, Madhu, and Hughes}]{Levitt02JMR}
\bibinfo{author}{M.~H. Levitt}, \bibinfo{author}{P.~Madhu},
  \bibinfo{author}{C.~E. Hughes},
\newblock \bibinfo{title}{Cogwheel phase cycling},
\newblock \bibinfo{journal}{J. Magn. Reson.} \bibinfo{volume}{155}
  (\bibinfo{year}{2002}) \bibinfo{pages}{300 -- 306}.
\bibitem[{Ernst et~al.(1987)Ernst, Bodenhausen, and Wokaun}]{Ernst87Book}
\bibinfo{author}{R.~R. Ernst}, \bibinfo{author}{G.~Bodenhausen},
  \bibinfo{author}{A.~Wokaun}, \bibinfo{title}{Principles of Nuclear Magnetic
  Resonance in One and Two Dimensions}, \bibinfo{publisher}{Clarendon Press},
  \bibinfo{address}{Oxford}, \bibinfo{year}{1987}.
\bibitem[{Ivchenko et~al.(2003)Ivchenko, Hughes, and Levitt}]{Ivchenko03JMR}
\bibinfo{author}{N.~Ivchenko}, \bibinfo{author}{C.~E. Hughes},
  \bibinfo{author}{M.~H. Levitt},
\newblock \bibinfo{title}{Multiplex phase cycling},
\newblock \bibinfo{journal}{J. Magn. Reson.} \bibinfo{volume}{160}
  (\bibinfo{year}{2003}) \bibinfo{pages}{52 -- 58}.
\bibitem[{Drobny et~al.(1978)Drobny, Pines, Sinton, Weitekamp, and
  Wemmer}]{Drobny78SympFaradSoc}
\bibinfo{author}{G.~Drobny}, \bibinfo{author}{A.~Pines},
  \bibinfo{author}{S.~Sinton}, \bibinfo{author}{D.~P. Weitekamp},
  \bibinfo{author}{D.~Wemmer},
\newblock \bibinfo{title}{Fourier transform multiple quantum nuclear magnetic
  resonance},
\newblock \bibinfo{journal}{Faraday Symp. Chem. Soc.} \bibinfo{volume}{13}
  (\bibinfo{year}{1978}) \bibinfo{pages}{49--55}.
\bibitem[{Walder et~al.(2013)Walder, Dey, Kaseman, Baltisberger, and
  Grandinetti}]{Walder13JCP}
\bibinfo{author}{B.~J. Walder}, \bibinfo{author}{K.~K. Dey},
  \bibinfo{author}{D.~C. Kaseman}, \bibinfo{author}{J.~H. Baltisberger},
  \bibinfo{author}{P.~J. Grandinetti},
\newblock \bibinfo{title}{Sideband separation experiments in {NMR} with phase
  incremented echo train acquisition},
\newblock \bibinfo{journal}{J. Chem. Phys.} \bibinfo{volume}{138}
  (\bibinfo{year}{2013}).
\bibitem[{Shannon(1949)}]{Shannon49ProcInstRadEng}
\bibinfo{author}{C.~E. Shannon},
\newblock \bibinfo{title}{Communication in the presence of noise},
\newblock \bibinfo{journal}{Proc. Institute of Radio Engineers}
  \bibinfo{volume}{37} (\bibinfo{year}{1949}) \bibinfo{pages}{10--12}.
\bibitem[{Grebenkov(2007)}]{Grebenkov07RevModPhys}
\bibinfo{author}{D.~S. Grebenkov},
\newblock \bibinfo{title}{{NMR} survey of reflected {Brownian} motion},
\newblock \bibinfo{journal}{Rev. Mod. Phys.} \bibinfo{volume}{79}
  (\bibinfo{year}{2007}) \bibinfo{pages}{1077--1137}.
\bibitem[{Stallmach and Galvosas(2007)}]{Stallmach07AnnuRepNMRSpectrosc}
\bibinfo{author}{F.~Stallmach}, \bibinfo{author}{P.~Galvosas},
\newblock \bibinfo{title}{Spin echo {NMR} diffusion studies},
\newblock in: \bibinfo{booktitle}{Annu. Rep. NMR Spectrosc.},
  volume~\bibinfo{volume}{61}, \bibinfo{publisher}{Elsevier, Amsterdam, Boston,
  Heidelberg}, \bibinfo{year}{2007}, pp. \bibinfo{pages}{51--131}.
\bibitem[{Holz et~al.(2000)Holz, Heil, and Sacco}]{Holz00PCCP}
\bibinfo{author}{M.~Holz}, \bibinfo{author}{S.~R. Heil},
  \bibinfo{author}{A.~Sacco},
\newblock \bibinfo{title}{Temperature dependent self-diffusion coefficients of
  water and six selected molecular liquids for calibration in accurate $^1${H}
  {NMR} {PFG}-measurements},
\newblock \bibinfo{journal}{Phys. Chem. Chem. Phys.} \bibinfo{volume}{2}
  (\bibinfo{year}{2000}) \bibinfo{pages}{4740}.
\bibitem[{Price and Kuchel(1991)}]{Price91JMR}
\bibinfo{author}{W.~S. Price}, \bibinfo{author}{P.~W. Kuchel},
\newblock \bibinfo{title}{Effect of nonrectangular field gradient pulses in the
  {S}tejskal and {T}anner (diffusion) pulse sequence},
\newblock \bibinfo{journal}{J. Magn. Reson.} \bibinfo{volume}{94}
  (\bibinfo{year}{1991}) \bibinfo{pages}{133--139}.
\bibitem[{Bluestein(1970)}]{Bluestein70IEEE}
\bibinfo{author}{L.~Bluestein},
\newblock \bibinfo{title}{A linear filtering approach to the computation of
  discrete fourier transform},
\newblock \bibinfo{journal}{IEEE Transactions on Audio and Electroacoustics}
  \bibinfo{volume}{18} (\bibinfo{year}{1970}) \bibinfo{pages}{451--455}.
\bibitem[{Hennig et~al.(1986)Hennig, Nauerth, and
  Friedburg}]{Hennig86MagnResonMed}
\bibinfo{author}{J.~Hennig}, \bibinfo{author}{A.~Nauerth},
  \bibinfo{author}{H.~Friedburg},
\newblock \bibinfo{title}{{RARE} imaging: {A} fast imaging method for clinical
  {MR}},
\newblock \bibinfo{journal}{Magn. Reson. Med.} \bibinfo{volume}{3}
  (\bibinfo{year}{1986}) \bibinfo{pages}{823--833}.
\bibitem[{Li et~al.(2009)Li, Han, and Balcom}]{Li09JMR_II}
\bibinfo{author}{L.~Li}, \bibinfo{author}{H.~Han}, \bibinfo{author}{B.~J.
  Balcom},
\newblock \bibinfo{title}{Spin echo {SPI} methods for quantitative analysis of
  fluids in porous media},
\newblock \bibinfo{journal}{J. Magn. Reson.} \bibinfo{volume}{198}
  (\bibinfo{year}{2009}) \bibinfo{pages}{252 -- 260}.
\bibitem[{Callaghan and Stepisnik(1995)}]{Callaghan95JMRComm}
\bibinfo{author}{P.~T. Callaghan}, \bibinfo{author}{J.~Stepisnik},
\newblock \bibinfo{title}{Frequency-domain analysis of spin motion using
  modulated-gradient {NMR}},
\newblock \bibinfo{journal}{J. Magn. Reson. A} \bibinfo{volume}{117}
  (\bibinfo{year}{1995}) \bibinfo{pages}{118--122}.
\bibitem[{Shemesh et~al.(2013)Shemesh, \'{A}lvarez, and Frydman}]{Shemesh13JMR}
\bibinfo{author}{N.~Shemesh}, \bibinfo{author}{G.~A. \'{A}lvarez},
  \bibinfo{author}{L.~Frydman},
\newblock \bibinfo{title}{Measuring small compartment dimensions by probing
  diffusion dynamics via non-uniform oscillating-gradient spin-echo ({NOGSE})
  {NMR}},
\newblock \bibinfo{journal}{J. Magn. Reson.} \bibinfo{volume}{237}
  (\bibinfo{year}{2013}) \bibinfo{pages}{49 -- 62}.

\end{thebibliography}

\end{document}